\documentclass[a4paper,11pt]{article}
\pdfoutput=1 

\usepackage{jcappub} 

\usepackage{lineno,multirow,xspace}\usepackage{microtype}
\usepackage[utf8]{inputenc}
\usepackage[numbers,sort&compress]{natbib}

\newcommand{\eV}{\mathrm{eV}}
\newcommand{\EeV}{\mathrm{EeV}}
\newcommand{\Mpc}{\mathrm{Mpc}}
\newcommand{\Rcut}{R_\text{cut}}
\newcommand{\sectionref}[1]{section~\ref{#1}}
\newcommand{\figureref}[1]{figure~\ref{#1}}
\newcommand{\SimProp}{\textit{SimProp}\xspace}
\newcommand{\dd}{\operatorname{d\!}}

\title{Secondary neutrino and gamma-ray fluxes from SimProp and CRPropa}

\author[a]{Rafael Alves~Batista,}
\author[b,c,d]{Denise Boncioli,}
\author[e]{Armando di~Matteo}
\author[d,f]{and Arjen van~Vliet}

\affiliation[a]{Instituto de Astronomia, Geofísica e Ciências Atmosféricas, Universidade de São Paulo, \\%
  Rua do Matão, 1226, 05508-090, São Paulo-SP, Brazil}
\affiliation[b]{Gran Sasso Science Institute (GSSI), Viale Francesco Crispi 7, 67100 L’Aquila, Italy}
\affiliation[c]{INFN, Laboratori Nazionali del Gran Sasso (LNGS), 67100 Assergi, L’Aquila, Italy}
  \affiliation[d]{Deutsches Elektronen-Synchrotron (DESY),\\%
	Platanenallee 6, 15738 Zeuthen, Germany}
\affiliation[e]{Service de Physique Théorique, CP255, Université Libre de Bruxelles (ULB),\\%
	Boulevard du Triomphe (Campus de la Plaine), 1050 Brussels, Belgium}
\affiliation[f]{Radboud University, Department of Astrophysics/IMAPP, \\%
	P.O. Box 9010, 6500 GL Nijmegen, The Netherlands}

\emailAdd{rafael.ab@usp.br}
\emailAdd{denise.boncioli@gssi.it}
\emailAdd{armando.di.matteo@ulb.ac.be}
\emailAdd{arjen.van.vliet@desy.de}

\abstract{

The interactions of ultra-high-energy cosmic rays (UHECRs) in extragalactic space with photons of the cosmic microwave background (CMB) and extragalactic background light (EBL) can generate high-energy neutrinos and photons. Simulations of UHECR propagation require knowledge about physical quantities such as the spectrum of the EBL and  photodisintegration cross sections. These assumptions, as well as the approximations used in the codes, may influence the computed UHECR spectrum and composition, and the associated cosmogenic neutrino and photon fluxes.  
Following up on our previous work where we studied the effects of these uncertainties on the UHECR spectrum and composition, here we quantify those on neutrino fluxes and production rates of photons, electrons, and positrons, using the Monte Carlo codes CRPropa and \SimProp, in various astrophysical scenarios. 
We show that cosmogenic neutrinos are more sensitive to the choice of EBL model than UHECRs, whereas the overall cosmogenic gamma-ray production rates are relatively independent of propagation details.  We also find significant differences between neutrino fluxes predicted by the latest released versions of CRPropa and \SimProp, and discuss their causes and possible improvements in future versions of the codes.
}

\begin{document}
	\maketitle
	\flushbottom
	\section{Introduction}\label{sec:introduction}

Ultra-high-energy cosmic rays (UHECRs) are widely believed to originate from extragalactic sources.  If this is the case, during their travel from their sources to Earth they are expected to interact with diffuse extragalactic background photons via photonuclear and photohadronic processes, losing energy and producing secondary particles, including neutrinos and photons. Similar processes may take place in photon fields in the vicinity of UHECR accelerator sites.

The photon backgrounds most relevant to UHECR propagation are the cosmic microwave background (CMB) and the extragalactic background light (EBL). 
The energy-loss lengths for cosmic-ray nuclei due to interactions with the CMB and EBL are shorter than the Hubble radius when nucleus energies~$E$ exceed a few~EeV ($1~\EeV \equiv 10^{18}~\eV$); those for protons with~$E \gtrsim 5 \times 10^{19}~\eV$ are a few hundred Mpc or less and those for composite nuclei with these energies even shorter, resulting in the well-known Greisen-Zatsepin-Kuz'min (GZK) cutoff \cite{Greisen:1966jv,Zatsepin:1966jv}.  These interactions can generate large amounts of cosmogenic photons and neutrinos.

Extragalactic cosmic rays are guaranteed to produce a flux of cosmogenic neutrinos with $E \sim 10$~PeV via interactions with EBL photons. If their maximum rigidity\footnote{The magnetic rigidity~$R$ of a particle is its momentum divided by its electric charge.  For ultrarelativistic fully ionized nuclei, $R \equiv E / Z$ in units where $c=e=1$.} is relatively high, i.e.~if the highest-energy cosmic rays are not dominated by heavy nuclei, then cosmogenic neutrinos with $E \sim 1$~EeV are also expected to be produced, via interactions with CMB photons.  Neutrinos are excellent cosmic messengers because they are not charged, hence their trajectories are unperturbed by magnetic fields, and due to their small cross sections they can propagate over cosmological distances bringing information from distant sources.  For example, different models of cosmological evolution of UHECR sources that would be hard to distinguish using cosmic-ray data alone can result in very different predictions of cosmogenic neutrino fluxes.
With the observation of extragalactic neutrinos by IceCube~\cite{Aartsen:2013jdh}, the interest in cosmogenic neutrinos has increased.  Nevertheless, only two neutrino events with~$E \gtrsim 2$~PeV have been detected so far \cite{Aartsen:2018vtx}, setting stringent upper bounds on cosmogenic neutrino fluxes. Next-generation neutrino detectors such as the Askaryan Radio Array (ARA)~\cite{Allison:2011wk}, the Antarctic Ross Ice-shelf ANtenna Neutrino Array (ARIANNA)~\cite{Glaser:2018ifj}, the Giant Radio Array for Neutrino Detection (GRAND)~\cite{Alvarez-Muniz:2018bhp} and the Probe Of Extreme Multi-Messenger Astrophysics (POEMMA)~\cite{Olinto:2017xbi} may be able to improve this picture.

Cosmogenic photons, too, are guaranteed to be produced via UHECR interactions, but they are expected to quickly interact with the CMB, EBL, and universal radio background (URB), resulting in electromagnetic cascades of gamma rays with energies up to a few hundred GeV, which can be observed with imaging air Cherenkov telescopes such as the upcoming Cherenkov Telescope Array (CTA)~\cite{Acharya:2017ttl}. Thus, cosmogenic photons are expected to contribute to the diffuse gamma-ray background (DGRB) observed at these energies, but fluxes of ultra-high-energy photons from extragalactic sources (except possibly very nearby ones) are expected to be extremely low, and none have been observed so far~\cite{Abu-Zayyad:2013dii,Aab:2016agp}.

The intrinsic synergies between these multiple cosmic messengers can be suitably and complementarily exploited to search for the elusive sources of UHECRs, whose origin, nature, and acceleration mechanisms are yet to be understood. 

Recently, there have been many works attempting to constrain the origin of UHECRs using neutrinos and/or photons~\cite{Hooper:2010ze,Murase:2012df,Heinze:2015hhp,Aloisio:2015ega,Supanitsky:2016gke,Globus:2017ehu,vanVliet:2019nse}.  Generally, these works simulate the propagation of UHECRs assuming specific models for the distribution and evolution of sources, and compute their cosmogenic fluxes.  By comparing these results with the available experimental data, it is possible to statistically constrain properties of UHECR sources, namely the spectrum and composition of the injected cosmic rays, the evolution and distribution of the sources over cosmological scales, and their overall luminosity.

Some of the physical quantities related to the production of cosmogenic neutrinos and photons are poorly known, such as the spectral energy density of the EBL at certain redshifts and wavelengths and the interaction cross sections for certain photodisintegration channels. Various models have been used to estimate these quantities in the aforementioned studies. Moreover, some of the computational approximations that have been used may have a non-negligible impact on the results. In the present paper we extend our previous work~\cite{Batista:2015mea} to quantify the effect of uncertainties related to the production of cosmogenic photons and neutrinos and compare two widely-used codes for UHECR propagation -- CRPropa~\cite{Batista:2016yrx} and \SimProp~\cite{Aloisio:2017iyh}.

This paper is structured as follows: in \sectionref{sec:interactions} we describe in detail the various processes in which cosmogenic neutrinos and photons can be produced; in \sectionref{sec:codes} we briefly describe the two simulation codes we used in this work; in \sectionref{sec:comparisons} we show our results about the effects of various possible settings for propagation simulations on computed predictions of various observable quantities; in \sectionref{sec:discussion} we discuss implications of our results and a few related issues; finally, in \sectionref{sec:conclusions} we briefly summarise our conclusions.


	\section{Photons and neutrinos from UHECRs}\label{sec:interactions}
\subsection{Production}
At ultra-high energies ($E \gtrsim 10^{18}~\eV$), the main interactions between cosmic rays and photon fields are the Bethe--Heitler electron--positron\footnote{Henceforth we refer to both electrons and positrons simply as `electrons', unless stated otherwise.} pair production, the photodisintegration of nuclei, and photopion production.  All these processes can directly or indirectly produce cosmogenic neutrinos\footnote{Throughout this work we collectively refer to both $\nu_l$ and $\bar\nu_l$ ($l = e, \mu, \tau$) simply as `neutrinos'.} and/or photons.

\newcommand{\nucl}[3]{{^{#1}_{#2}\mathrm{#3}}}
\newcommand{\gbg}{\gamma_\text{bg}}
\newcommand{\gHE}{\gamma_\text{HE}}
For protons with $E \lesssim 60~\EeV$, the dominant process is Bethe--Heitler pair production: $\nucl{A}{Z}{X} + \gbg \rightarrow \nucl{A}{Z}{X} + e^+ + e^-$, where $\gbg$ denotes the background photon. 
This process has a very small inelasticity ($\lesssim 0.1\%$), but a very short mean free path ($< 1~\Mpc$ for protons with $E\gtrsim 6~\EeV$), so that over cosmological distances it can result in sizeable energy losses and the production of large numbers of PeV-energy electrons.
These can subsequently initiate electromagnetic cascades as mentioned in \sectionref{sec:neutralpropa}, contributing to the cosmogenic gamma-ray flux.

For composite nuclei with $E \lesssim 40~\EeV$~per nucleon, the dominant process is photodisintegration, whereby a nucleus is stripped of one or more nucleons or other light fragments such as deuterons or alpha particles, via photonuclear interactions.  The dominant channel is single-nucleon ejection.  Notations such as $\nucl{A}{Z}{X}(\gamma,p)\nucl{A-1}{Z-1}{X'}$ for $\nucl{A}{Z}{X} + \gbg \to \nucl{A-1}{Z-1}{X'} + p$, etc., are sometimes used.
Typical interaction lengths for this process are $\sim 50$~Mpc for nuclei with energies $\sim 1~\EeV$~per nucleon (mainly on EBL photons), and $\lesssim 1$~Mpc for energies $\gtrsim 10~\EeV$~per nucleon (mainly on CMB photons).
The residual nucleus and the ejected fragments can be unstable to beta decay, producing electrons and neutrinos, or to isomeric transitions, producing gamma rays, but generally these are less energetic and have lower fluxes than neutrinos, electrons, and photons stemming from photopion production.
Cross sections for photodisintegration are not totally well-known, which may affect secondary particle yields, as discussed in \sectionref{sec:disint_comp}. 
Another process of interest is elastic scattering of background photons by cosmic rays ($\nucl{A}{Z}{X} + \gbg \to \nucl{A}{Z}{X} + \gHE$), which can increase the total photon flux, though this contribution is small.

At the highest energies, the dominant process is photopion production, which can affect both free nucleons and ones bound within nuclei (which are thereby ejected from them).
The channels with the lowest thresholds are $p + \gbg \to p + \pi^0$ and $p + \gbg \to n + \pi^+$,
via the $\Delta^+$ resonance. The pion produced has $\sim 20\%$ of the initial nucleon energy on average;
it quickly decays as $\pi^0 \to 2\gHE$ or $\pi^+ \to \mu^+ + \nu_\mu$, $\mu^+ \to e^+ + \bar\nu_\mu + \nu_e$.  Each photon produced in this way has $\sim 10\%$ of the initial nucleon's energy, and each neutrino or electron $\sim 5\%$.
At higher energies, the production of multiple pions, and/or heavier hadrons, becomes possible.
The energy loss length for pion production on CMB photons is shorter than the Hubble radius for $E \gtrsim 50$~EeV/nucleon, and sharply decreases at higher energies, down to $\sim 15$~Mpc; this is expected to be the dominant source of cosmogenic neutrinos with $E \sim 1$~EeV. 
Pion production on EBL photons has an energy loss length longer than the Hubble radius for every value of $E$/nucleon, but it is still expected to be the dominant source of cosmogenic neutrinos at $E \sim 10$~PeV.

On their way to Earth, UHECRs can be deflected by extragalactic (EGMF) and Galactic (GMF) magnetic fields, altering their arrival direction distribution. Furthermore, EGMFs can increase the length of their trajectories (especially at lower energies) and hence the expected number of interactions, resulting in harder UHECR spectra and larger fluxes of secondaries.  EGMFs are poorly known and models attempting to describe them vary largely; for a comparison and detailed discussion see Ref.~\cite{AlvesBatista:2017vob}.  On the other hand, these effects are only significant if the average spacing between sources is comparable to or larger than the energy loss length and/or the Larmor radius; otherwise, the propagation theorem \cite{Aloisio:2004jda} predicts that the results of the propagation are independent of magnetic deflections.
Throughout this work we only consider homogeneous source distributions, to which the propagation theorem is applicable. Moreover, only one-dimensional propagation in the absence of magnetic fields is implemented in \SimProp. Therefore, the study of the effects of EGMF uncertainties is outside the scope of this work.

\subsection{Propagation}\label{sec:neutralpropa}

Once produced, neutrinos and antineutrinos travel in straight lines, with negligible probability of interacting with any other particles in the intergalactic space. The only processes affecting them are adiabatic energy losses due to the expansion of the
Universe (a neutrino produced at redshift $z$ with energy $E$ reaches us with energy $E/(1+z)$) and flavour oscillations.

Conversely, electrons and photons quickly initiate electromagnetic cascades via repeated interactions with the diffuse extragalactic background radiation, such as pair production ($\gHE + \gbg \rightarrow e^+ + e^-$), inverse Compton scattering ($e^\pm + \gbg \rightarrow e^\pm +  \gHE$), double pair production ($\gHE + \gbg \rightarrow 2e^+ + 2e^-$), and triplet pair production ($e^\pm + \gbg \rightarrow e^\pm + e^+ + e^-$).
The electrons and photons reaching Earth will have energies $\lesssim 100$~GeV, with spectra largely independent of the original energy of the electrons or photons that initiated the cascade (provided it is $\gtrsim 100$~TeV) and only weakly dependent on the redshift of its production point~\cite{Berezinsky:2016feh}. Hence, to characterise the cosmogenic gamma-ray fluxes produced in a given scenario, only the total energy of the electrons and photons produced at each redshift is needed, not the energy distribution of the individual electrons and photons.

Intergalactic magnetic fields\footnote{Here we distinguish between extragalactic (EGMF) and intergalactic magnetic fields (IGMFs). The latter refers to the magnetic fields in cosmic voids, whereas the former generally describes any magnetic field outside our Galaxy, including those in filaments and clusters of galaxies.} (IGMFs) can cause the charged leptonic component of the cascade to lose energy via synchrotron emission; for example, a 10 EeV electron loses $0.3$ EeV/Mpc in a magnetic field of 0.1 nG~\cite{Heiter:2017cev}.
IGMFs may also deflect electrons and positrons in the cascade resulting in a broadening of the arrival directions of the order of $B/(10^{-14}~\mathrm{G})$~degrees \cite{Aharonian:1993vz,Neronov:2009gh}, suppressing the gamma-ray flux coming directly from the sources and contributing to the diffuse gamma-ray background. The non-observation of the so-called ``pair haloes''~\cite{Neronov:2010bi,Ackermann:2013yma,Biteau:2018tmv,Broderick:2018nqf} implies that strong magnetic fields ($B \gtrsim 3\times 10^{-13}$~G)  are required to fully isotropise the cascade and explain the absence of haloes (c.f.~Ref.~\cite{Chen:2014rsa}). An alternative explanation~\cite{Broderick:2018nqf} suggests that plasma instabilities could cause electrons to quickly cool, dumping energy into the intergalactic medium and consequently quenching the cascade. Another possible propagation effect is the oscillation of photons into hypothetical axion-like particles in the presence of IGMFs~\cite{Horns:2012kw,Meyer:2013pny}.

	\section{UHECR propagation simulation codes}\label{sec:codes}

\subsection{CRPropa}

CRPropa\footnote{\url{https://crpropa.desy.de}}~\cite{Armengaud:2006fx,Kampert:2012fi,Batista:2016yrx} is a framework designed for simulations of UHECR propagation in the galactic and extragalactic spaces. It also includes the production and propagation of secondary electrons, neutrinos, and photons. CRPropa 3~\cite{Batista:2016yrx} is written in C++ with Python bindings. Its modular design provides an intuitive way to assemble functionalities to enable a wide range of studies. Recent extensions of the code include additional photon production channels~\cite{Heiter:2017cev} and a low-energy extension for galactic cosmic-ray propagation~\cite{Merten:2017mgk}.

The main differences between CRPropa and \emph{SimProp} were described in Ref.~\cite{Batista:2015mea}. Therefore, we refer the reader to this previous work for more details on the implementation of interactions. One change of the current version of CRPropa with respect to the one used in our previous work is the change in the default photodisintegration cross sections for nuclei with $A \ge 12$, from TALYS 1.6 to TALYS 1.8~\cite{Koning:2012zqy}, though for most channels relevant to UHECR propagation the differences between them are negligible.
Cross sections for nuclei with $A \le 12$ are the same as in CRPropa 2 (see~\cite{Kampert:2012fi} and references therein for more details; note that the cross sections for the photodisintegration of helium nuclei were recently shown to be overestimated~\cite{Soriano:2018lly}).
Here we describe how secondary photons, electrons, and neutrinos are produced in CRPropa.

The dominant channel for secondary neutrino production at ultra-high energies is pion decay ($\pi^+ \to \mu^+ + \nu_\mu$, $\mu^+ \to e^+ + \bar\nu_\mu + \nu_e$), produced via photopion production. In CRPropa this process is modelled with the SOPHIA package~\cite{Mucke:1999yb}. The branching ratio of pion production determines the proton-to-neutron ratio that in turn determines the ratio between neutrinos and photons produced. While this quantity can be obtained by simple isospin symmetry considerations, a full Monte Carlo treatment is preferred. Another important quantity is the multiplicity of pion production, which will directly impact the proton-to-neutron ratio depending on the energy fraction carried by each pion. Note that the branching ratio is energy-dependent, and at the threshold near the $\Delta^+$ resonance the charged pion channel dominates over the neutral one~\cite{Mucke:1999yb}.

The most important process for the production of secondary electrons and positrons is Bethe-Heitler pair production ($\nucl{A}{Z}{X} + \gbg \rightarrow \nucl{A}{Z}{X} + e^+ + e^-$). Because of the very small inelasticity and very short mean free path of this process it is treated in CRPropa using a continuous energy loss approximation.

Photodisintegration processes may render the final state of a nucleus excited; its subsequent decay can emit photons, typically with energies $\sim 10^{15} \; \text{eV}$ for $\sim 10^{19} \; \text{eV}$ cosmic rays. The branching ratios are given by the ratio between the cross sections of the mother and its corresponding excited daughter nucleus ($\sigma_{i \rightarrow f^\text{*}} ^\text{PD}$), divided by the photodisintegration cross section for a mother nucleus to disintegrate into an specific daughter nucleus ($\sigma_{i \rightarrow f} ^\text{PD}$), i.e., $\sigma_{i \rightarrow f^\text{*}} ^\text{PD} / \sigma_{i \rightarrow f} ^\text{PD}$, wherein $i$ and $f$ denotes, respectively, the mother and daughter nuclei, and the asterisk refers to the excited state.

Unstable nuclei can decay and produce multiple photons during gamma decay (see Sec.~\ref{sec:interactions}). The multiplicity of the photons as well as their energy are obtained from the NuDat 2.6 database\footnote{National Nuclear Data Center, {\url{http://www.nndc.bnl.gov/nudat2/}}.}. Elastic scattering (${^A _Z X} + \gbg \rightarrow {^A _Z X} + \gHE$) is also implemented in CRPropa. More details about the implementation of additional photon channels in CRPropa are described in Ref.~\cite{Heiter:2017cev}.

In this work we use the latest version, CRPropa~3~\cite{Batista:2016yrx}, which is continuously being maintained and improved and is publicly available from {\url{https://crpropa.desy.de}}.

In CRPropa the redshift evolution of the EBL background is obtained using a global scaling factor, $s(z)$. This quantity is obtained by dividing the integrated comoving spectral number density of EBL photons at redshift $z$ by the corresponding value at $z = 0$, as follows:
\begin{equation}
	s(z) \equiv \dfrac{\int^\infty_0 n(\epsilon, z) \dd\epsilon}{\int^\infty_0 n(\epsilon, 0) \dd\epsilon},
\end{equation}
where $n(\epsilon,z)$ is the number of photons per unit comoving volume and per unit photon energy~$\epsilon$.
Therefore, the mean free path~$\lambda$ for a given interaction scales as~\cite{Batista:2016yrx}
\begin{equation}
	\lambda(E, z) = \dfrac{\lambda(E (1 + z), 0)}{(1 + z)^3 s(z)}.
	\label{eq:lambda}
\end{equation}
In addition, neutrino flavour oscillations are not implemented, but those can be trivially taken into account downstream in the data analysis.

	\subsection{\SimProp}
\SimProp is a fast and simple UHECR propagation code, first released in 2011 \cite{Aloisio:2012wj}
in order to have a publicly available Monte Carlo code for the community to use
at a time when most UHECR propagation studies used closed-source simulation codes.
More sophisticated codes such as CRPropa have since become available,
but \SimProp remains useful due to its computational speed and ease of use,
and since it uses different algorithms and approximations,
it can provide independent cross-checks of simulation results.
For example, the importance of the previously neglected dependence
of the UHECR spectrum and composition at Earth in certain injection scenarios
on EBL and (to a lesser extent) photodisintegration models \cite{ArmandoCombFit,Boncioli:2015pds,Aab:2016zth}
was first discovered during an investigation on the origin of differences
between results of early \SimProp and CRPropa versions \cite{Batista:2015mea},
as were major differences between photodisintegration cross sections computed by
preliminary \cite{Khan:2004nd} and released \cite{Goriely:2008zu} versions of TALYS.

The first \SimProp version \cite{Aloisio:2012wj} was developed in \cite{Boncioli:2014hrt}
as a refinement of the analytic models by Aloisio, Berezinsky and Grigorieva
\cite{Aloisio:2008pp,Aloisio:2010he}.
The only processes treated stochastically in it were the injection and
the photodisintegration of nuclei.  Adiabatic energy losses, electron--positron
pair photoproduction, and pion photoproduction were treated deterministically
according to the continuous energy loss approximation, as in the analytic models
\SimProp was based on.

In subsequent versions, \SimProp was extended with the stochastic treatment
(and tracking of secondary particles) of pion production by CMB photons \cite{Aloisio:2013kea},
then of pion production by EBL photons \cite{Aloisio:2015sga,Aloisio:2015ega},
then photodisintegration with alpha-particle ejection \cite{Aloisio:2016tqp,diMatteo:2017ndg},
then electron--positron pair production \cite{Aloisio:2017iyh}.
New versions also extended the range of user options
(e.g. choice of photodisintegration and EBL modeling and of output formats),
fixed bugs, and improved the calculation speed of the previous versions.

In this work we use the latest version, \SimProp~v2r4 \cite{Aloisio:2017iyh},
released in May 2017 and available upon request to
\newcommand{\mail}[1]{\href{mailto:#1}{\nolinkurl{#1}}}
\mail{SimProp-dev@aquila.infn.it}.  Several EBL models are available in it, among which the best-fit model from Dom\'inguez et~al.\ (2011) \cite{Dominguez:2010bv} and the fiducial model from Gilmore et~al.\ (2012) \cite{Gilmore:2011ks}.
As for photodisintegration, PSB cross sections \cite{Puget:1976nz,Stecker:1998ib} are used for $A \le 4$
(which were recently shown to considerably overestimate the photodisintegration rate of helium \cite{Soriano:2018lly}),
whereas for heavier nuclei the user can choose either the PSB model or Gaussian or Breit-Wigner cross sections with arbitrary parameter values,
and parameter files fitted to cross sections computed by TALYS \cite{Khan:2004nd,Goriely:2008zu,Koning:2012zqy} are distributed with \SimProp.

Approximations used in \SimProp include:
treating all photohadronic interactions as single-pion production with branching ratios from isospin invariance
($2/3$ neutral pions, $1/3$ charged pions)
and distributing the direction of the momentum of the outgoing pion isotropically in the centre-of-mass frame;
treating all particle decays as instantaneous, as decay times are usually much shorter than all other relevant timescales
except for a few beta decays;
a simplified treatment of nuclear photodisintegration;
and computing the number of electron--positron pairs produced assuming that all production occurs at the threshold
(since as mentioned in \sectionref{sec:neutralpropa} only their total energy is actually relevant).
Conversely, unlike in CRPropa, for most of the available EBL models the EBL spectral densities are interpolated as a function of both
the redshift and the photon energy from 2D grids of values adapted from the original works,
without any assumption about the redshift evolution of the spectral shape.

Interactions of electrons and photons are not implemented in \SimProp yet:
only the redshifts of their production points and their initial energies are written in the output,
so that the resulting electromagnetic cascades can be simulated using external programs.
Likewise, and as in CRPropa, neutrino flavour oscillations are not implemented,
but can be trivially taken into account downstream in the data analysis.
Unlike in CRPropa, in the current version of \SimProp magnetic deflections are not implemented: all particle trajectories are treated as one-dimensional, the redshift~$z$ being the only coordinate kept track of.

	\newcommand{\lgE}{\log_{10}(E/\mathrm{eV})}
\section{Comparisons}\label{sec:comparisons}


In our comparisons, we consider source scenarios assuming injection spectra of the form
\begin{equation}
	\frac{\dd N_\text{inj}}{\dd E} \propto \left(\frac{E}{\mathrm{EeV}}\right)^{-\gamma} \exp\left(-\frac{E}{Z\Rcut}\right),
\end{equation}
where the source emissivity as a function of the redshift~$z$ is expressed as
\begin{align}
	\mathcal{L}(z) = \int_{E_{\min}}^{E_{\max}} \frac{ E\dd N_\text{inj}}{\dd E \dd V_\text{c} \dd t} \dd E &= \mathcal{L}_0 S(z), &\mathcal{L}(0) &= \mathcal{L}_0,\quad S(0) = 1,\end{align}
where $\dd V_\text{c} = (1+z)^3\dd V$ is the comoving volume element.
The various combinations of parameters we used are listed in Table~\ref{tab:inj}.
In all cases we took $E_{\min}=10^{16}$~eV, $E_{\max}=10^{23}$~eV, $z_{\min}=0$ and $z_{\max}=6$.
The first three scenarios are based on the ``dip model'' scenarios from Ref.~\cite{Aloisio:2015ega}, but without the spectral break at low energies. The last four were used in Ref.~\cite{Batista:2015mea}, and are loosely inspired by the ``best fit'' and the ``second local minimum'' scenarios of Ref.~\cite{Aab:2016zth}.
Because we assume a continuous distribution of sources, it follows from the propagation theorem~\cite{Aloisio:2004jda} that the effects of magnetic deflections can be neglected when computing the fluxes. This justifies our one-dimensional treatment.  Discrete source distributions are outside the scope of this work.

\begin{table}[t]
	\centering
	\caption{The various injection scenarios we considered}
	\label{tab:inj}
	\begin{tabular}{r@{~}lr@{}lcccc}
		\hline\hline
		\multicolumn{2}{c}{\multirow{2}{*}{name}} & \multirow{2}{*}{$^A$}&\multirow{2}{*}{$Z$} & $\mathcal{L}_0$ & \multirow{2}{*}{$S(z)$} & \multirow{2}{*}{$\gamma$} & $\Rcut$ \\
		\cline{5-5} \cline{8-8}
		~ & ~ & ~ & ~ & {\footnotesize $\mathrm{erg}~\mathrm{Mpc}^{-3}~\mathrm{yr}^{-1}$} & ~ & ~ & {\footnotesize $\mathrm{EV}$} \\
		\hline
		unif.&protons & $^{1}$&H & $15\phantom{.0}\times{10}^{45}$ & 1 & $2.6$ & $10^4$ \\
		SFR&protons & $^{1}$&H & $\phantom{0}6\phantom{.0}\times{10}^{45}$ & $S_\text{SFR}(z)$ & $2.5$ & $10^4$ \\
		AGN&protons & $^{1}$&H & $\phantom{0}3.5\times{10}^{45}$ & $S_\text{AGN}(z)$ & $2.4$ & $10^4$ \\
		hard&nitrogen & $^{14}$&N & $\phantom{0}0.5\times{10}^{45}$ & 1 & $1\phantom{.0}$ & $\phantom{00}5$ \\
		hard&iron & $^{56}$&Fe & $\phantom{0}0.5\times{10}^{45}$ & 1 & $1\phantom{.0}$ & $\phantom{00}5$ \\
		soft&nitrogen & $^{14}$&N & $\phantom{0}1\phantom{.0}\times{10}^{45}$ & 1 & $2\phantom{.0}$ & $100$ \\
		soft&iron & $^{56}$&Fe & $\phantom{0}1\phantom{.0}\times{10}^{45}$ & 1 & $2\phantom{.0}$ & $100$ \\
		\hline
	\end{tabular}
	\small\begin{align*}
		S_\text{SFR}(z) &= \begin{cases}
			(1+z)^{3.4}, & 0 \le z\le 1; \\
			2^{3.7}(1+z)^{-0.3}, & 1 \le z \le 4; \\
			2^{3.7}5^{3.2}(1+z)^{-3.5}, & 4 \le z \le 6.
		\end{cases}& S_\text{AGN}(z) &= \begin{cases}
			(1+z)^5, & 0.0 \le z\le 1.7; \\
			2.7^5, & 1.7 \le z \le 2.7; \\
			2.7^5 10^{2.7-z}, & 2.7 \le z \le 6.0.
		\end{cases}
	\end{align*}
\end{table}

We performed a series of comparisons between simulation results to assess the effects of different cosmological parameters, different EBL models, different photodisintegration models, and different propagation codes.  For each pair of propagation models, we simulated UHECR propagation for each of the injection scenarios listed in table~\ref{tab:inj} and computed cosmic-ray fluxes at Earth from~$\lgE = 17.5$ to~$20.5$ in bins of width~0.1, and cosmogenic neutrino fluxes from~$\lgE = 15$ to~$20$. In the cases of nitrogen and iron injection, we also computed $\langle\ln A\rangle$ and $\sigma^2_{\ln A}$ in the same energy bins as for the spectra.

In the case of electrons and photons, their propagation is not implemented in \SimProp and it is outside the scope of this paper, so we only compare their production rates, in redshift bins $[10^{-4.0}, 10^{-3.8})$, $\ldots$, $[10^{+0.4}, 10^{+0.6})$, $[10^{+0.6}, 6 \approx 10^{0.78})$. For each bin, we compute the total energy of the photons and electrons produced in it, divided by $(1+z)$ and by the width of the bin in light-travel distance so that the total energy flux reaching Earth in the form of cascades is the integral of this quantity.

Note that, whereas the proton injection scenarios can approximately reproduce the observed total UHECR spectrum down to $\sim 1~\EeV$,
the nitrogen or iron injection scenarios can only do so above $\sim 5~\EeV$ (the so-called ankle in the UHECR spectrum) and drastically underestimate it at lower energies.
Any realistic scenario must include an additional population making up the majority of cosmic rays below the ankle,
whose mass composition appears to be proton-dominated and whose nearly isotropic angular distribution suggests an extragalactic origin~\cite{Abbasi:2016kgr}, which can be presumed to result in cosmogenic neutrino fluxes $\lesssim 250$~PeV similar to those in proton-only scenarios, as well as considerable cosmogenic gamma-ray fluxes~\cite{Liu:2016brs}.
Therefore, in the case of our nitrogen and iron scenarios, comparisons involving nuclei with $E \lesssim 10^{18.7}~\eV$, neutrinos with $E \lesssim 10^{17.4}~\eV$ and gamma rays are for diagnostic purposes only, and not directly relevant to predictions in realistic scenarios, where the contribution of the additional proton population would dominate.  In the corresponding plots, these energy ranges are indicated by a shaded background.
Also,  since cascades initiated by electrons and photons are not distinguishable at Earth, only the sum is relevant for predicting observations; nevertheless, we compute electron and photon production rates separately because they have different sensitivities to assumptions and approximations implemented in the algorithms.
Cascades initiated by photons are subdominant with respect to those initiated by electrons by an order of magnitude in proton and soft injection scenarios, and by two orders of magnitude in hard injection scenarios, so even sizeable differences in photon production rates have only a minor impact on the total cascade density. We shade the background of plots of photon production rates to indicate this fact.

Where not otherwise noted, we used the \SimProp simulation code, $H_0 = 70$~km/s/Mpc, $\Omega_\text{m} = 0.3$, $\Omega_\Lambda = 0.7$,
the Gilmore et~al.~2012 fiducial \cite{Gilmore:2011ks} EBL model,
and photodisintegration cross sections computed using TALYS~\cite{Goriely:2008zu} with parameters restored to those of the preliminary version~\cite{Khan:2004nd} as described in Ref.~\cite{Batista:2015mea}.

\subsection{Effects of different cosmological parameters}
\label{sec:cosmo_comp}

\begin{figure}
	\centering
	\includegraphics[width=3in]{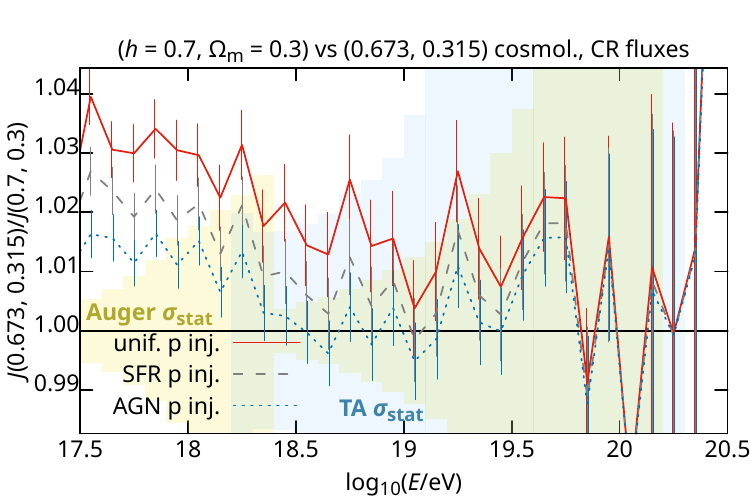}
	\includegraphics[width=3in]{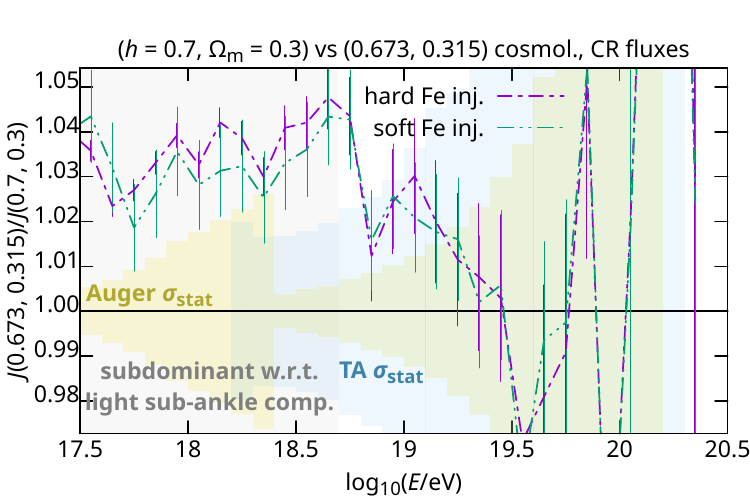}%
	\caption{
		Ratios between predicted UHECR fluxes, using CRPropa, in various scenarios assuming two different sets of cosmological parameters.
		Results in nitrogen injection scenarios are very similar to those for iron, and are not shown.
		Statistical uncertainties of Pierre Auger Observatory~\cite{Augerspectrum} and Telescope Array~\cite{TAspectrum} measurements are also shown for comparison.
	}
	\label{fig:cosmo-ncomp}
\end{figure}

\begin{figure}
	\centering
	\includegraphics[width=3in]{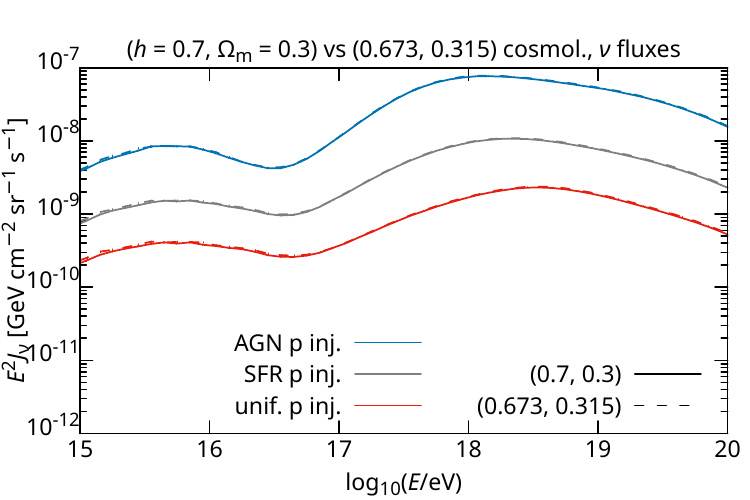}
	\includegraphics[width=3in]{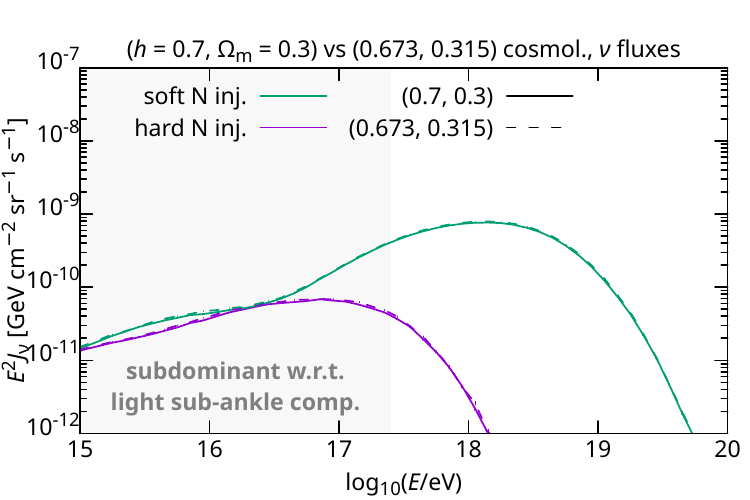}\\
	\includegraphics[width=3in]{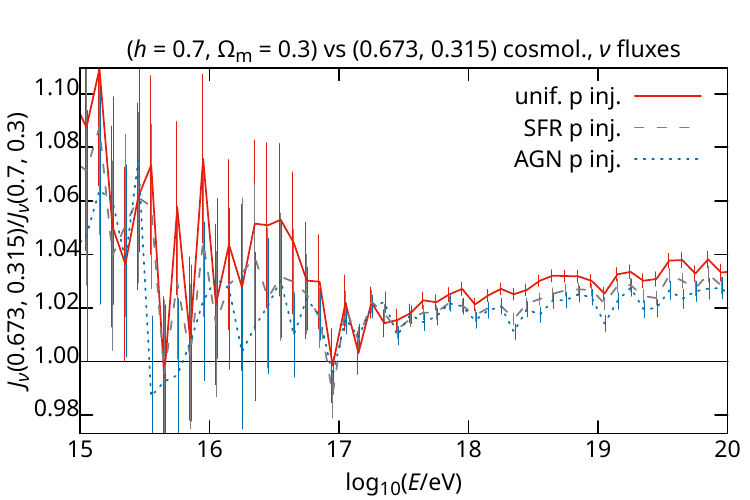}
	\includegraphics[width=3in]{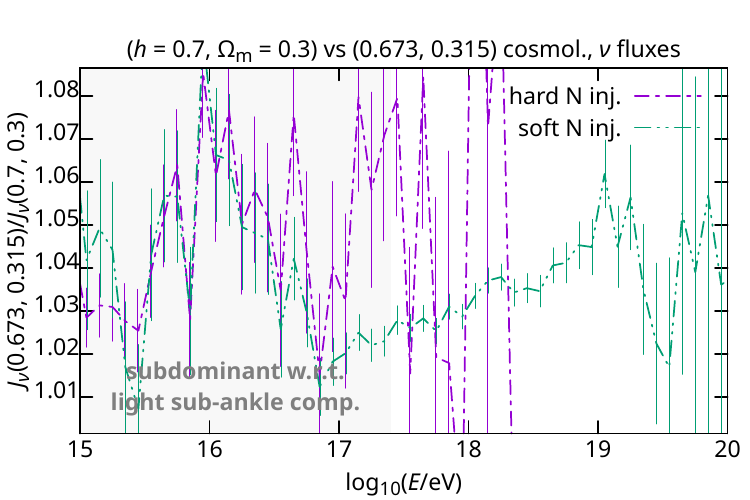}%
	\caption{
		Predicted cosmogenic neutrino fluxes, using CRPropa, in various scenarios assuming two different sets of cosmological parameters,
		and their ratios.
		Results in iron injection scenarios are very similar to those for nitrogen, and are not shown.
	}
	\label{fig:cosmo-nucomp}
\end{figure}

\begin{figure}
	\centering
	\includegraphics[width=3in]{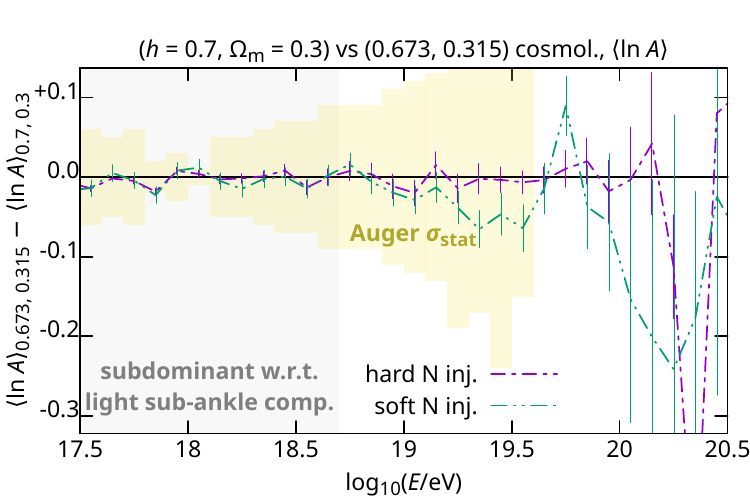}
	\includegraphics[width=3in]{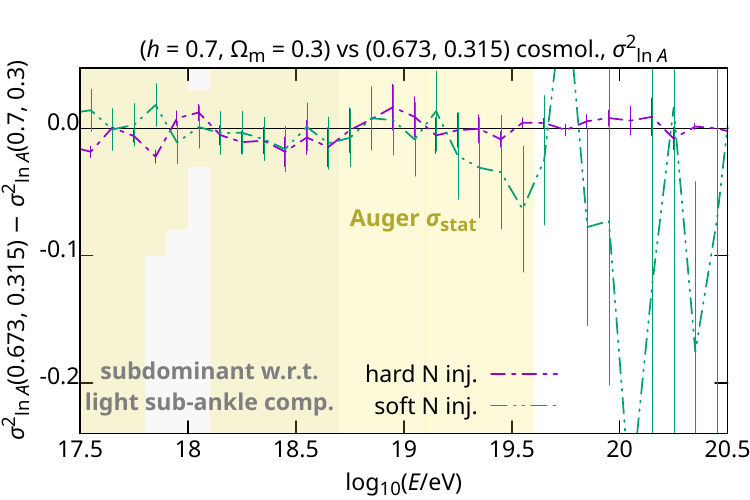}%
	\caption{Differences between the first two moments of the predicted UHECR mass composition, using CRPropa, in two nitrogen injection scenarios assuming two different sets of cosmological parameters.
	Differences in iron injection scenarios are even smaller and are not shown.
	Statistical uncertainties of Pierre Auger Observatory measurements \cite{Augercomp} are also shown for comparison.}
	\label{fig:cosmo-lnAcomp}
\end{figure}

The values of cosmological parameters used in \SimProp are $H_0 = 70$~km/s/Mpc, $\Omega_\text{m} = 0.3$, and $\Omega_\Lambda = 0.7$,
whereas those used by default in CRPropa are $H_0 = 67.3$~km/s/Mpc, $\Omega_\text{m} = 0.315$, and $\Omega_\Lambda = 0.685$.
These differences are comparable to the current experimental uncertainties.\footnote{%
The latest results from the Planck collaboration \cite{Aghanim:2018eyx} are
$H_0 = (67.4 \pm 0.5)$~km/s/Mpc, $\Omega_\text{m} = 0.315 \pm 0.007$, and $\Omega_\Lambda = 0.685 \pm 0.007$.  The agreement with the CRPropa values is coincidental, because when the CRPropa values were chosen, the latest Planck values then available \cite{Ade:2015xua} were roughly halfway between \SimProp and CRPropa values.}
To check the effect of these differences, we compared results of CRPropa simulations with both triples of parameters.
We found that the latter values result in slightly steeper (spectral index increase $\sim 0.005$) predicted UHECR spectra (\figureref{fig:cosmo-ncomp}),  a few per~cent larger UHECR and neutrino fluxes (figures \ref{fig:cosmo-ncomp} and \ref{fig:cosmo-nucomp}), and no discernible difference in the UHECR mass compositions (\figureref{fig:cosmo-lnAcomp}).
Since the overall normalization of fluxes is normally used as a free parameter, there being no direct knowledge of source emissivities at this level of precision, and fitting uncertainties on the UHECR spectral index are usually of order $\pm 0.1$ or larger, this means that the common practice of neglecting the uncertainties on cosmological parameters in UHECR propagation studies, sometimes adopting rounded values for them, is fully justified.

\subsection{Effects of different EBL models}\label{sec:EBL_comp}

Direct observations of the EBL are rather hard due to the presence of a much brighter foreground, the zodiacal light.
Various phenomenological models have been developed to describe its energy spectrum and cosmological evolution, e.g.~
\cite{Kneiske:2003tx,Stecker:2005qs,Stecker:2006eh,Franceschini:2008tp,Dominguez:2010bv,Gilmore:2011ks,Stecker:2012ta,Scully:2014wpa,Stecker:2016fsg,Franceschini:2017iwq,Andrews:2017ima,Ajello:2018}, based on a variety of techniques including semi-analytical modelling, observations, and/or a combination of the two.  Until recently, there were large discrepancies among them at all wavelengths. However, thanks to gamma-ray observations, the most recent models~\cite{Dominguez:2010bv,Gilmore:2011ks} are in good agreement in the near-infrared, visible, and ultraviolet wavelengths at $z=0$, with very narrow uncertainty bands.
On the other hand, at far-infrared wavelengths there remain large differences (over a factor of 2)
between models even at $z=0$. These were shown to have considerable impact in predictions of the UHECR spectrum and composition at Earth~\cite{Batista:2015mea}, which in turn were shown to result in major differences in the injection parameters required to fit the observed UHECR spectrum and composition data~\cite{Aab:2016zth}.
At higher redshifts,  extrapolations often have to be made, generating even more discrepancies among models. Contrary to the case of UHECRs, these may significantly affect neutrino predictions, as the contribution of distant sources to the total flux is important.

To assess the effect of these uncertainties, we compared propagation results using two of the most recent EBL models: Gilmore et~al.~2012~\cite{Gilmore:2011ks} fiducial (G12) and Dom\'inguez et~al.~2011~\cite{Dominguez:2010bv} best-fit (D11).  At $z=0$, D11 has about twice the spectral energy density of G12 in the far-infrared but about 10\% less than G12 in the visible/ultraviolet; the latter difference is larger at high redshifts.
As shown in Ref.~\cite{Batista:2015mea}, in proton injection scenarios D11 results in a $3\%$~lower flux at $E \approx 40$~EeV due to increased pion production on far-infrared photons; this difference is smaller than the statistical uncertainties of current UHECR spectrum measurements in bins of width $\Delta\lgE = 0.1$.  For nuclei injection, the differences are much larger, of the order of $30\%$ for hard injection and $10\%$ for soft injection. D11 results in fewer surviving nuclei around 40~EeV and more secondary protons around 10~EeV, effectively softening the observed spectrum at Earth, thus requiring a harder injection spectrum to reproduce a given observed spectrum at Earth~\cite{Aab:2016zth}.  The increased photodisintegration with D11 also results in a lighter mass composition at Earth.

\begin{figure}
	\centering
	\includegraphics[width=3in]{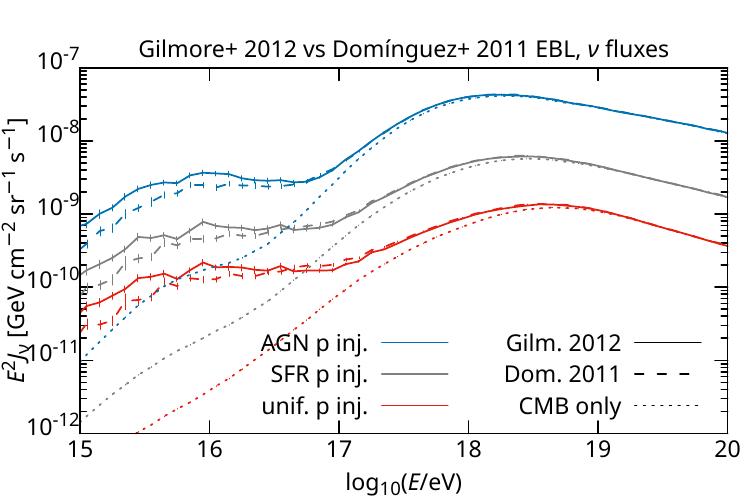}
	\includegraphics[width=3in]{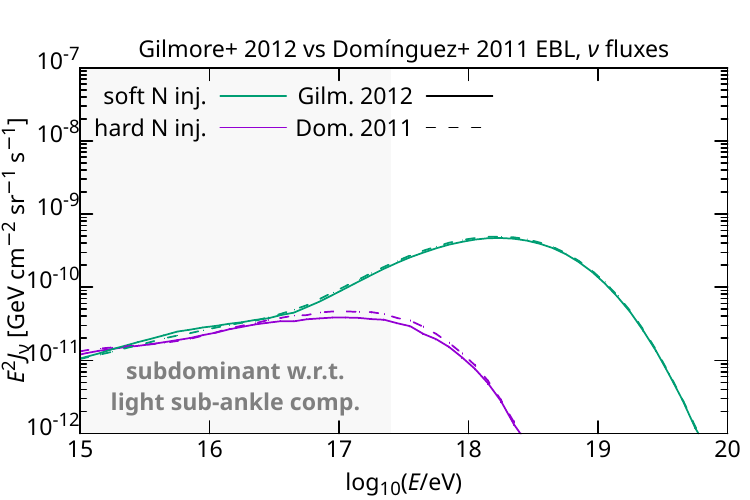}\\
	\includegraphics[width=3in]{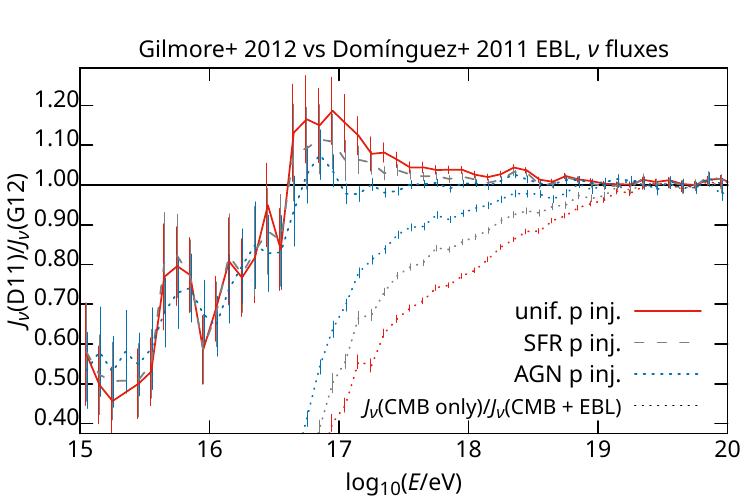}
	\includegraphics[width=3in]{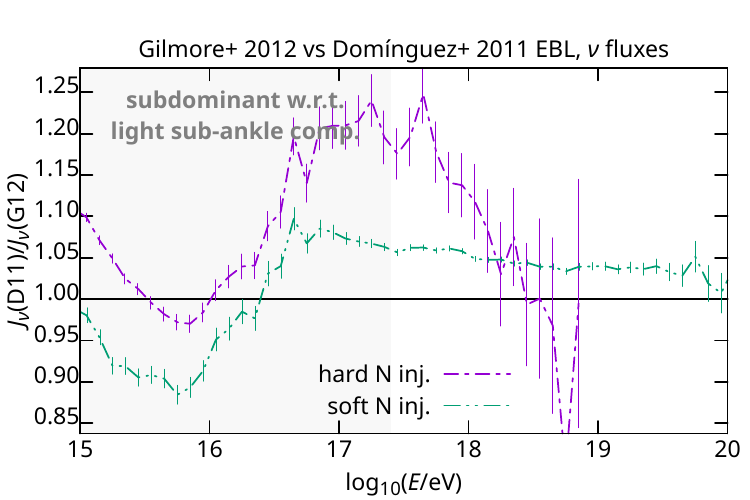}%
	\caption{
		Predicted cosmogenic neutrino fluxes, using \SimProp, in various scenarios assuming two different EBL models,
		and their ratios. Results in iron injection scenarios are very similar to those for nitrogen, and are not shown.
	}
	\label{fig:EBLs-nucomp}
\end{figure}

\begin{figure}
	\centering
	\includegraphics[width=3in]{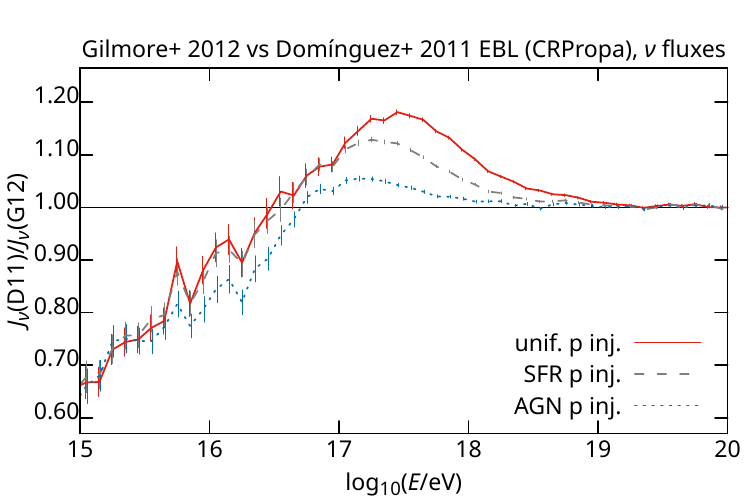}
	\includegraphics[width=3in]{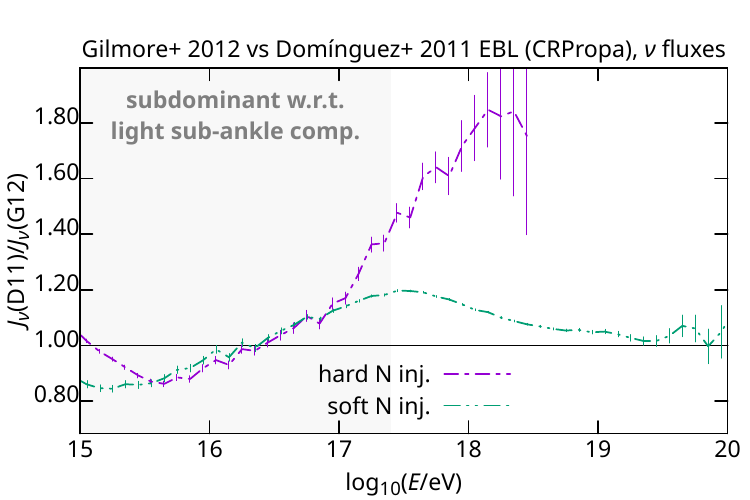}%
	\caption{
		Same as \figureref{fig:EBLs-nucomp}, bottom panels, but using CRPropa simulations.
	}
	\label{fig:CRP_EBLs-nucomp}
\end{figure}

In the left panels of \figureref{fig:EBLs-nucomp} we show that in proton injection scenarios neutrino fluxes are about the same at the highest energies (where pion production on the CMB dominates) for both EBL models; at intermediate energies (where pion production on far-infrared photons dominates) they are slightly higher with D11 than with G12, and vice-versa at the lowest energies (where pion production on visible/ultraviolet photons dominates). A similar behaviour is found for nitrogen injection, as shown in the right panels of \figureref{fig:EBLs-nucomp}; in this case, the energy below which photopion production on visible/ultraviolet photons dominates over that on far-infrared photons is lowered by an order of magnitude because of the reduced energy per nucleon, and at the lowest energies the difference between the overall neutrino fluxes is partially compensated (or even reversed, for hard nitrogen injection) by the increased numbers of neutrinos from the beta decay of products of photodisintegration on far-infrared photons in the D11 EBL model.  In \figureref{fig:CRP_EBLs-nucomp}, we show the analogous comparisons using CRPropa simulations, showing qualitatively similar results but with somewhat larger differences at higher energies and smaller differences at low energies, due to the approximation used in CRPropa for the EBL time evolution, which results in larger differences at high~$z$ between the two EBL models in the far IR and smaller differences in the near~IV/visible/UV.

\begin{figure}
	\centering
	\includegraphics[width=3in]{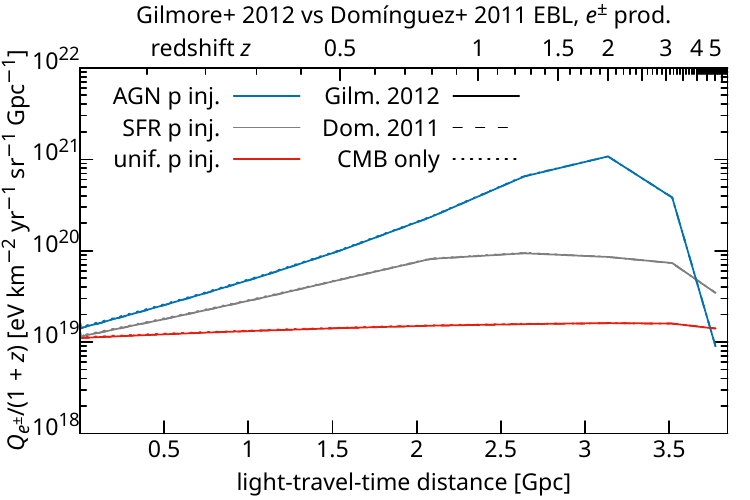}
	\includegraphics[width=3in]{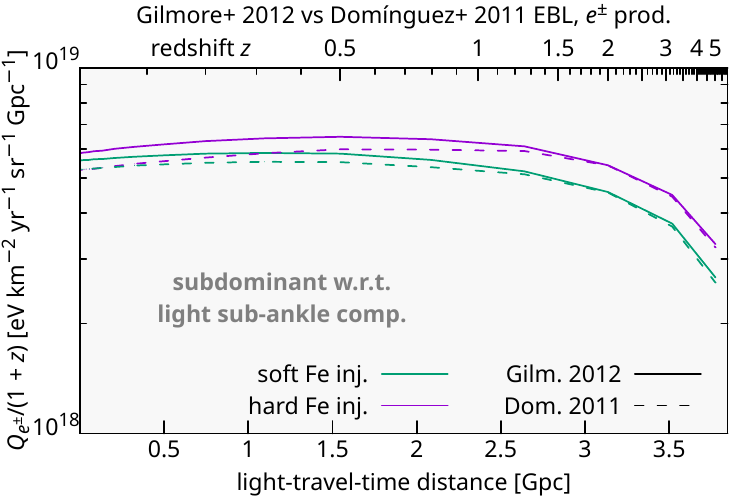}\\
	\includegraphics[width=3in]{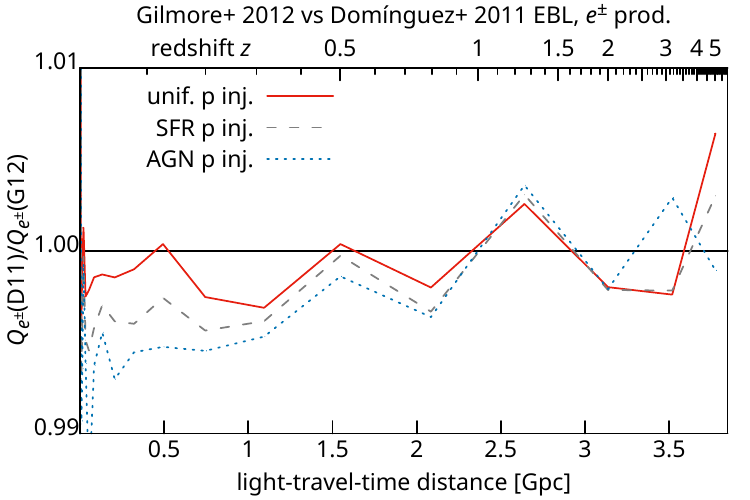}
	\includegraphics[width=3in]{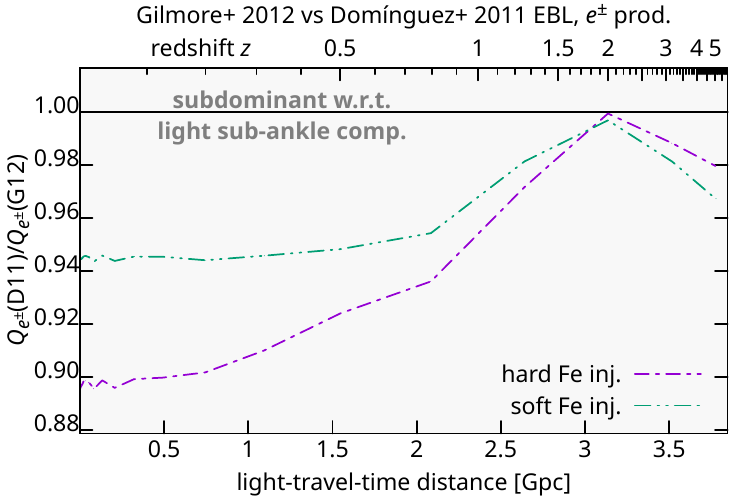}\\
	\includegraphics[width=3in]{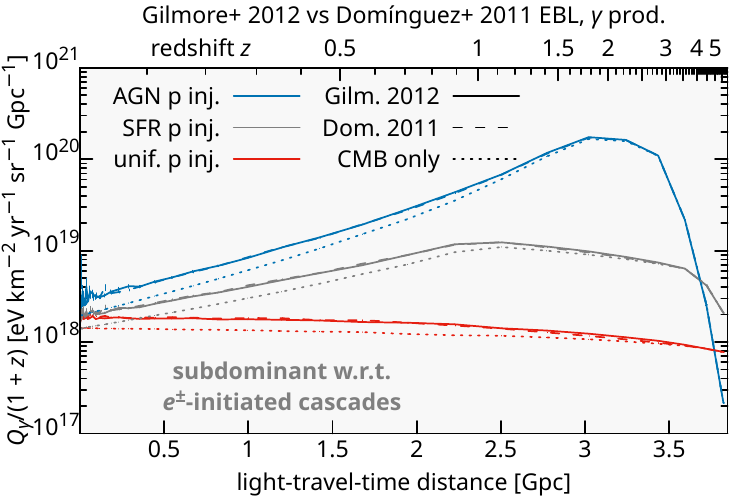}
	\includegraphics[width=3in]{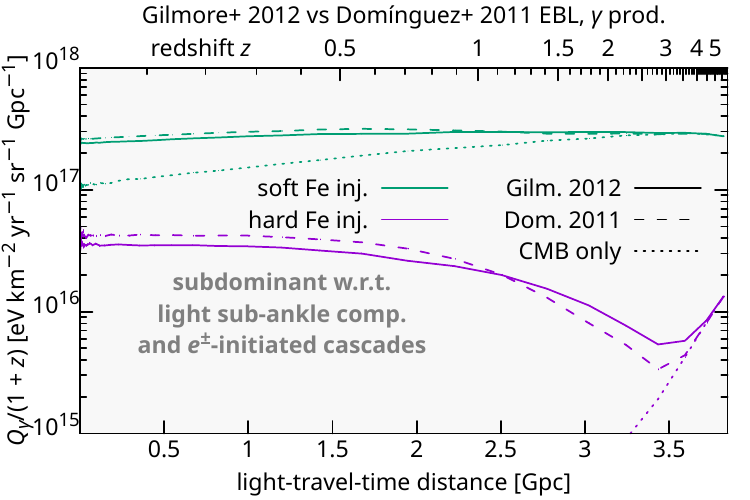}\\
	\includegraphics[width=3in]{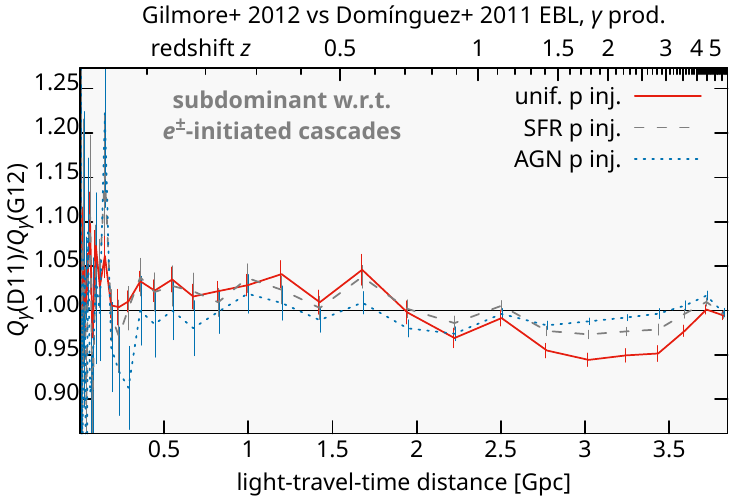}
	\includegraphics[width=3in]{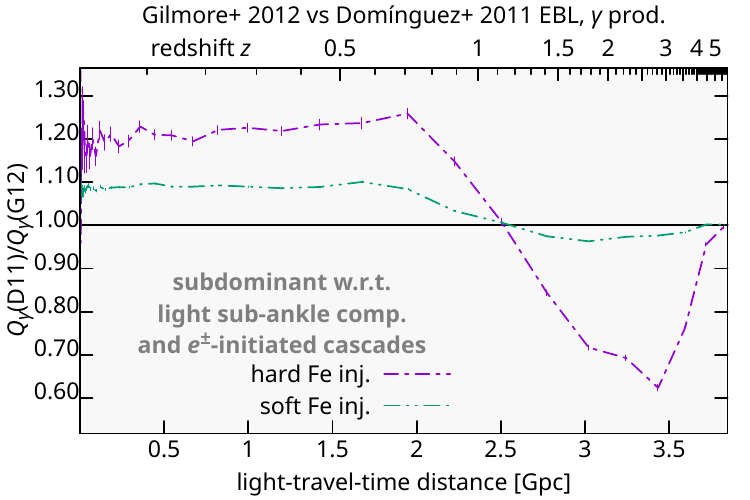}%
	\caption{Predicted cosmogenic $e^\pm$ and $\gamma$ production rates, using \SimProp, for two EBL models, and their ratios.
	Differences for nitrogen injection are slightly smaller than for iron and not shown.}
	\label{fig:EBLs-gammacomp}
\end{figure}

As shown in \figureref{fig:EBLs-gammacomp}, the differences between the production rates of electromagnetic cascades initiated by electrons and positrons for the two EBL models are of the order of a few per~mille for proton injection, and a few per~cent for nuclei injection.  The differences between production rates of cascades initiated by photons are somewhat larger, but as mentioned before, these are indistinguishable from and subdominant with respect to those initiated by electrons and positrons.  (The reason for the non-monotonic redshift dependence of the photon production rate in the hard iron injection scenario is that with such a low rigidity cutoff, pion production on the CMB is only substantial at~$z \gtrsim 3$ and that on the EBL only at $z \lesssim 3$.)
%
\subsection{Effects of different photodisintegration models}\label{sec:disint_comp}

Total photoabsorption cross sections have been measured for only 14 nuclides among nuclear isotopes interesting for cosmic-ray astrophysics~\cite{Batista:2015mea,Boncioli:2016lkt,Soriano:2018lly}. The experimental situation becomes better if we consider the inclusive cross sections for the emission of neutrons. Conversely, exclusive cross sections for channels in which all ejected fragments are charged are harder to measure. In particular, ${^{A}_{Z}X}(\gamma, \alpha){^{A-4}_{Z-2}X'}$ channels have only been measured for a handful of nuclides. These have a noticeably different impact on UHECR propagation than single-nucleon ejection as the secondaries have four times as much energy.

Phenomenological models to estimate such cross sections are available, but not necessarily reliable. For instance, TALYS overestimates the exclusive cross section of ${^{12}\mathrm{C}}(\gamma, \alpha){^{8}\mathrm{Be}}$ by an order of magnitude (see the Appendix A of Ref.~\cite{Batista:2015mea} and references therein), whereas the PSB model \cite{Puget:1976nz,Stecker:1998ib} neglects such channels altogether.  To assess the effects of the uncertainty in the cross sections, we used \SimProp to compare results obtained with TALYS and PSB cross sections for $A>4$. For $A \le 4$, PSB cross sections were used in both cases. Note that these were recently shown to overestimate the photodisintegration rate of helium~\cite{Soriano:2018lly}.

As already shown in Ref.~\cite{Batista:2015mea}, the main difference in the CR fluxes is the much higher flux of secondary helium with TALYS compared to PSB, which is explained by the fact that the former model includes channels that directly produce secondary alpha particles, whereas in the latter they are only produced after a nucleus loses single nucleons all the way down to ${^8\mathrm{Be}}(\gamma,n)2\alpha$.
In the case of iron injection, secondary helium is strongly subdominant so that its effect on the total fluxes is minor, whereas for nitrogen injection, the differences in the overall fluxes are comparable to those from different EBL models.

\begin{figure}
	\centering
	\includegraphics[width=3in]{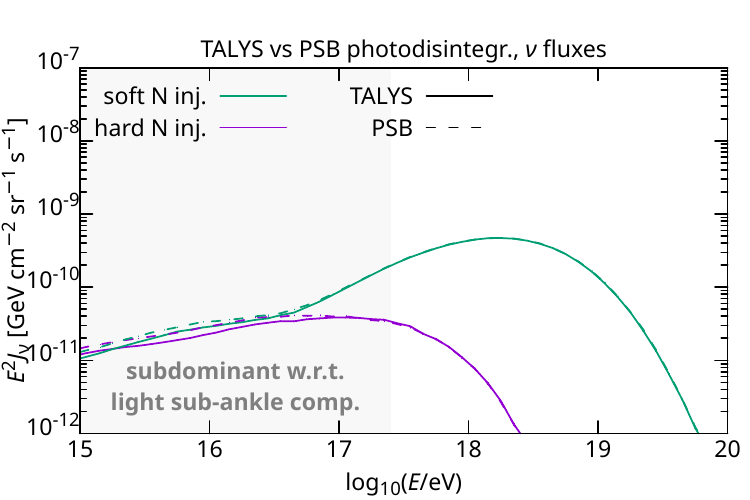}
	\includegraphics[width=3in]{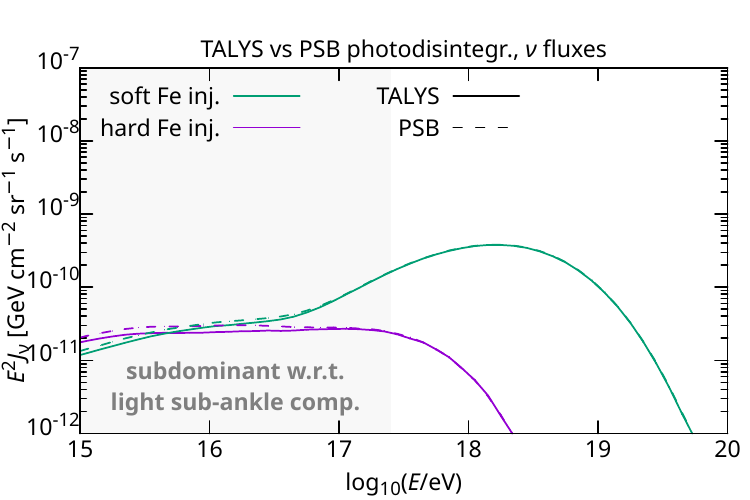}\\
	\includegraphics[width=3in]{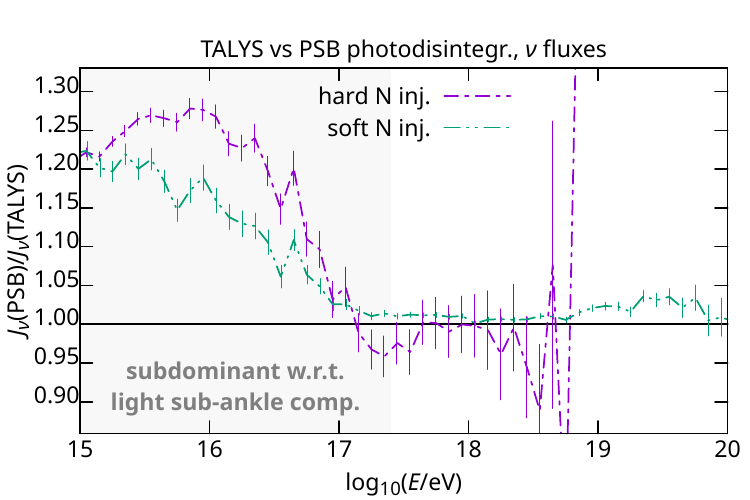}
	\includegraphics[width=3in]{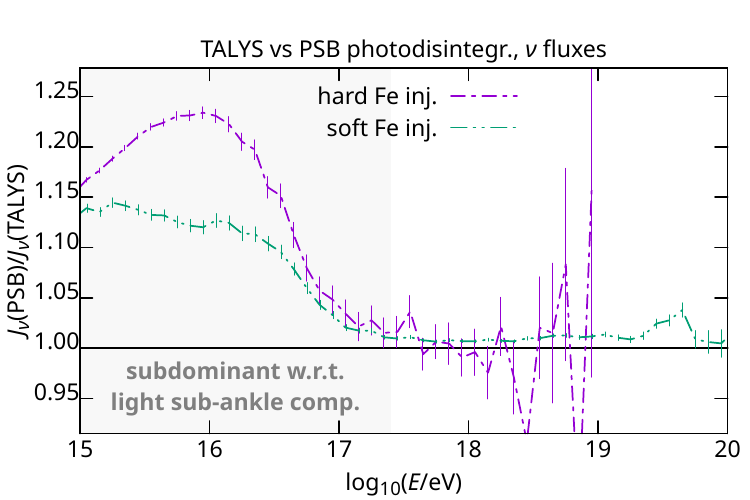}%
	\caption{
		Predicted cosmogenic neutrino fluxes, using \SimProp, in various scenarios assuming two different photodisintegration models,
		and their ratios
	}
	\label{fig:sigma-nucomp}
\end{figure}

In \figureref{fig:sigma-nucomp} we show that different photodisintegration models cause no sizeable differences in neutrino fluxes at EeV energies, for either nitrogen or iron injection. In fact, at these energies neutrinos result mainly from photopion production, which affects both free nucleons and those bound in nuclei in approximately the same way, and consequently are not affected by photodisintegration. On the other hand, below about 100~PeV, the neutrino flux is dominated by beta decay (in these artificial scenarios with no light sub-ankle UHECR component).  In the PSB photodisintegration model as implemented in \SimProp, only free nucleons are ejected, their types being randomly chosen, consequently overestimating the number of beta decays as, in reality, interaction channels yielding stable nuclei are more likely. This effect is weaker with the TALYS photodisintegration model, because it also includes channels ejecting fragments like $\alpha$ particles, reducing the number of beta-unstable nuclei produced.  This effect is stronger in the hard injection scenarios, because in soft injection scenarios, due to the higher injection cutoff, pion production is more frequent, so that the contribution of beta-decay neutrinos to the overall flux is smaller.

\begin{figure}
	\centering
	\includegraphics[width=3in]{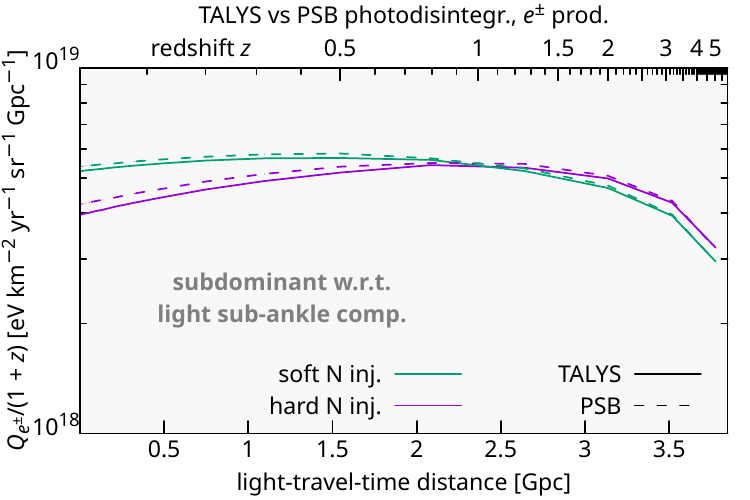}
	\includegraphics[width=3in]{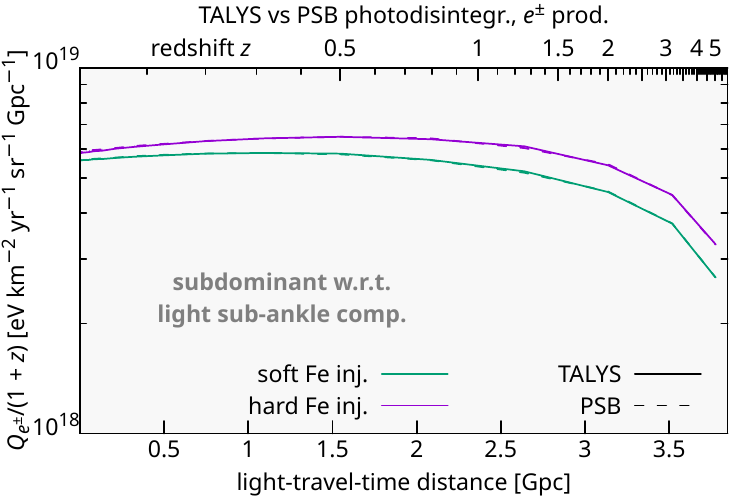}\\
	\includegraphics[width=3in]{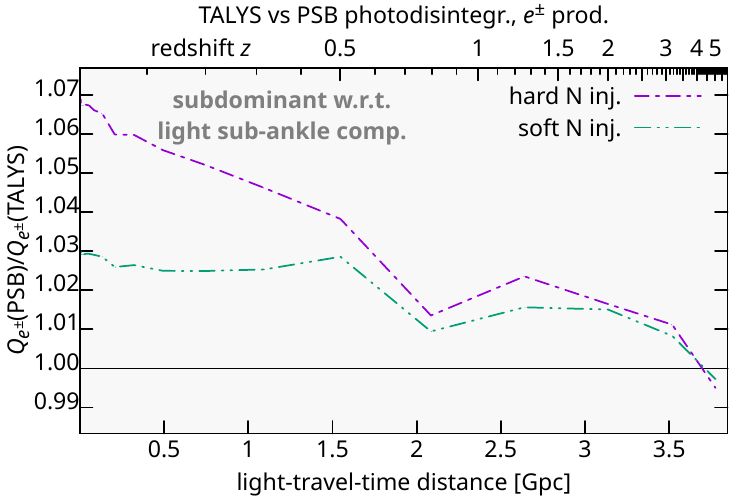}
	\includegraphics[width=3in]{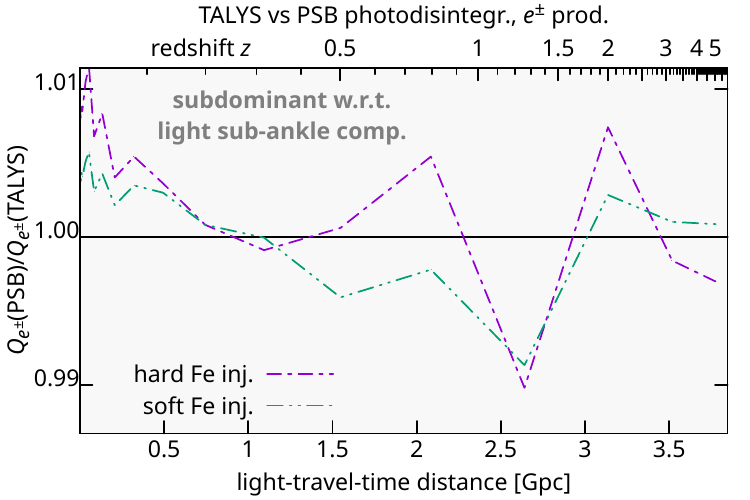}\\
	\includegraphics[width=3in]{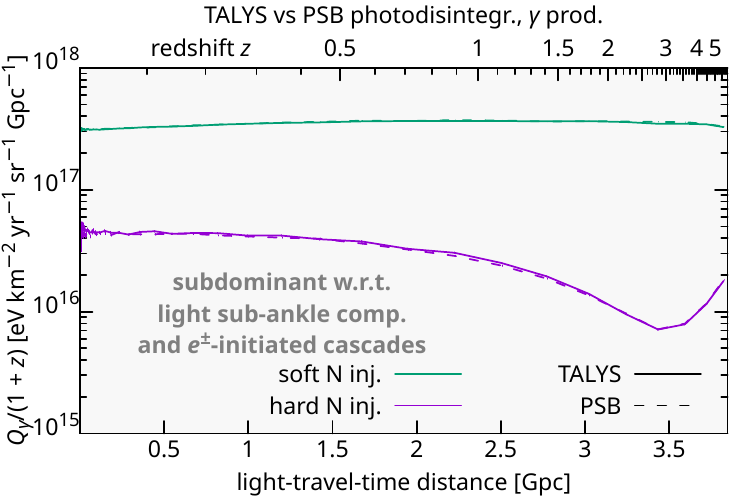}
	\includegraphics[width=3in]{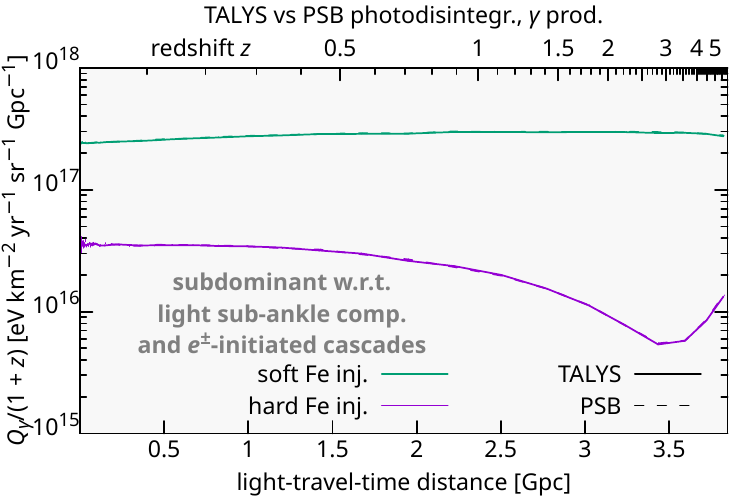}\\
	\includegraphics[width=3in]{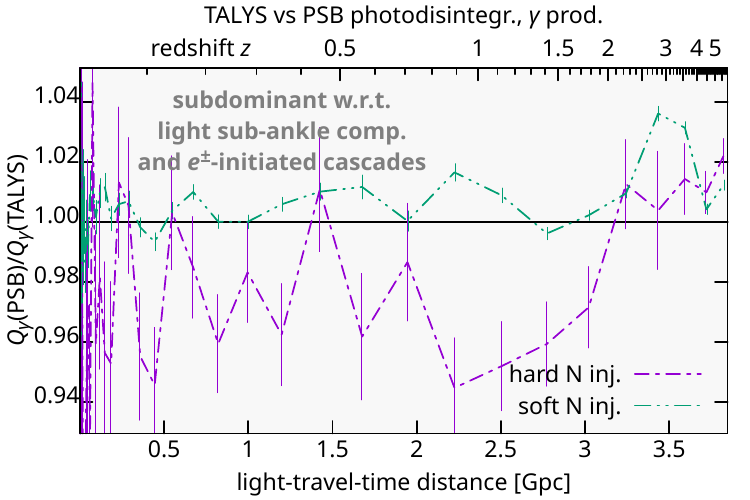}
	\includegraphics[width=3in]{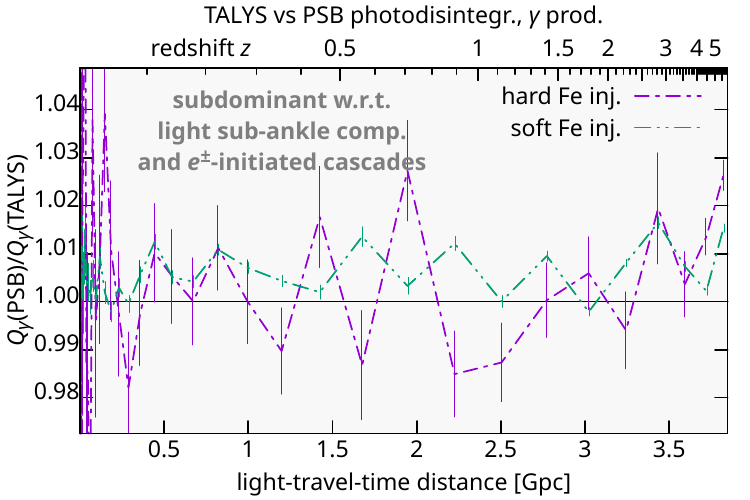}%
	\caption{Predicted cosmogenic $e^\pm$ and $\gamma$ production rates, using \SimProp, in various scenarios assuming two different photodisintegration models,
		and their ratios.}
	\label{fig:sigma-gammacomp}
\end{figure}

Similarly, in \figureref{fig:sigma-gammacomp} we show that the effects of the differences between the two photodisintegration models on predicted $e^\pm$ and $\gamma$ production rates are minor, of the order of a few per~cent or less for nitrogen injection and a few per~mille or less for iron injection.

\subsection{Effects of different propagation codes}\label{sec:codes_comp}
\begin{figure}
	\centering
	\includegraphics[width=3in]{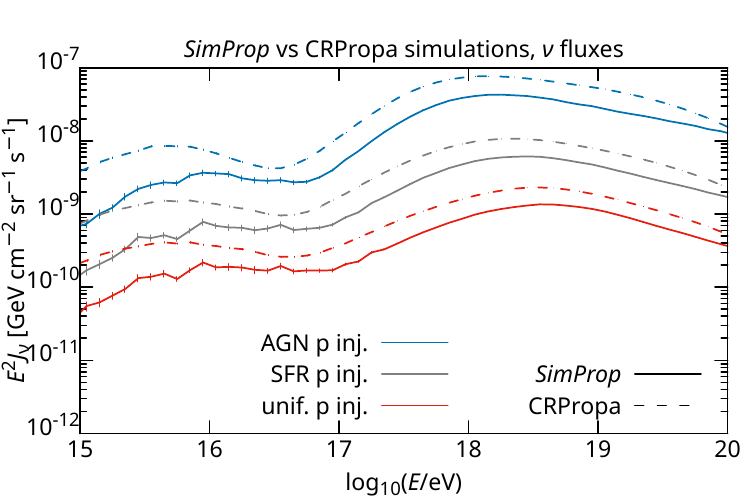}
	\includegraphics[width=3in]{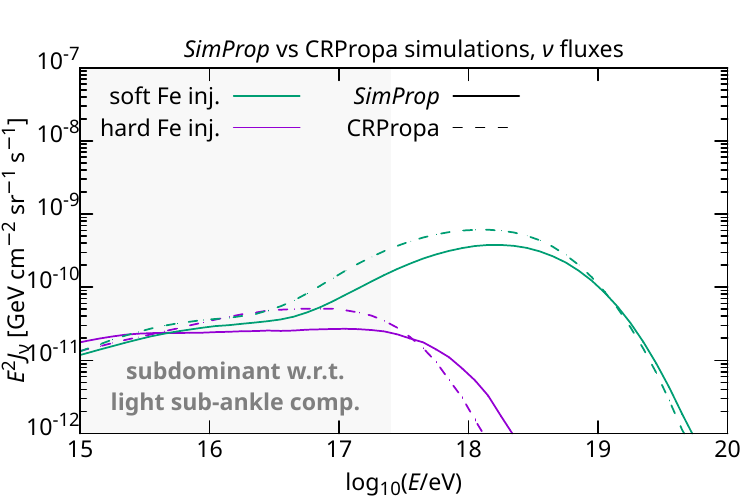}\\
	\includegraphics[width=3in]{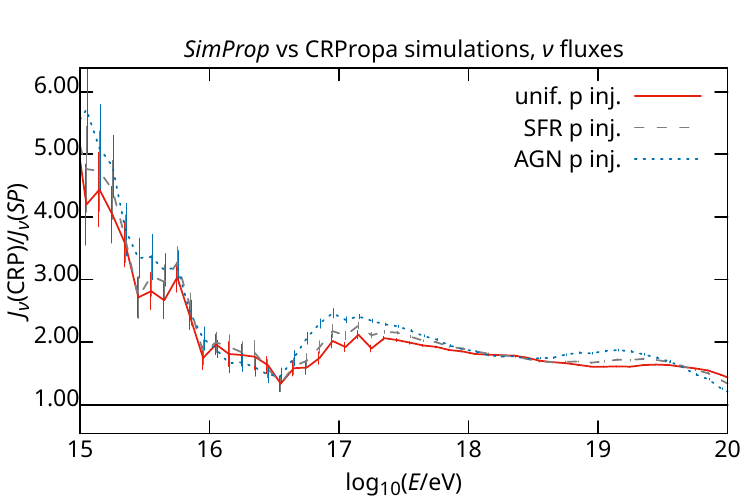}
	\includegraphics[width=3in]{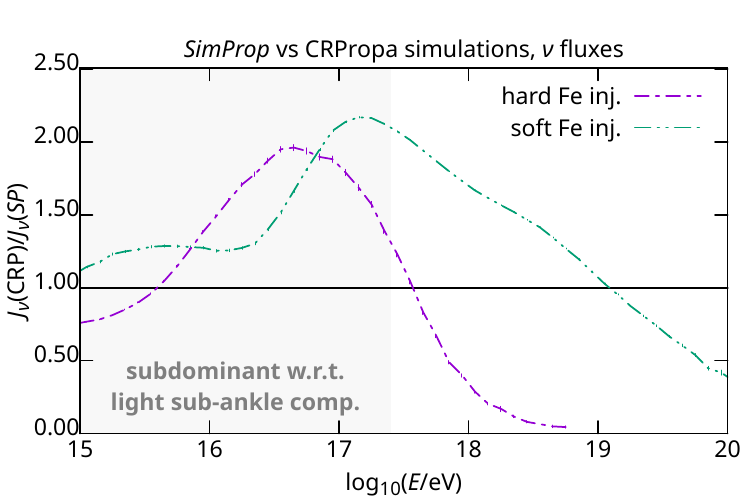}%
	\caption{
		Cosmogenic neutrino fluxes in various scenarios predicted by two different simulation codes,
		and their ratios.
		Results in nitrogen injection scenarios are very similar to those for iron, and are not shown.
	}
	\label{fig:codes-nucomp}
\end{figure}

\begin{figure}
	\centering
	\includegraphics[width=3in]{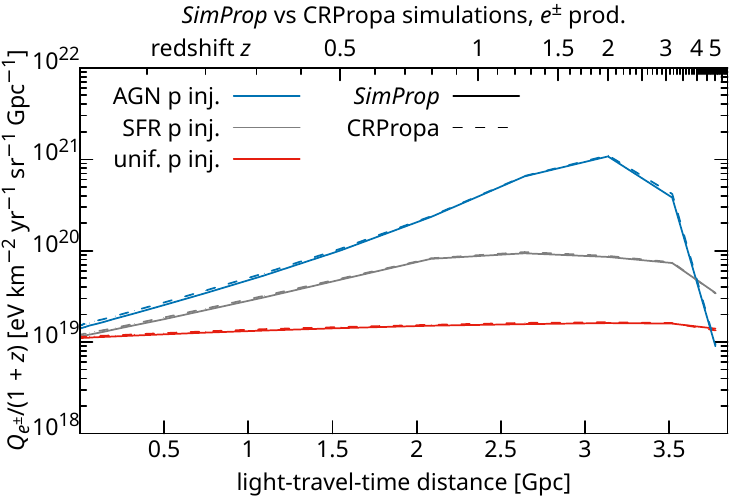}
	\includegraphics[width=3in]{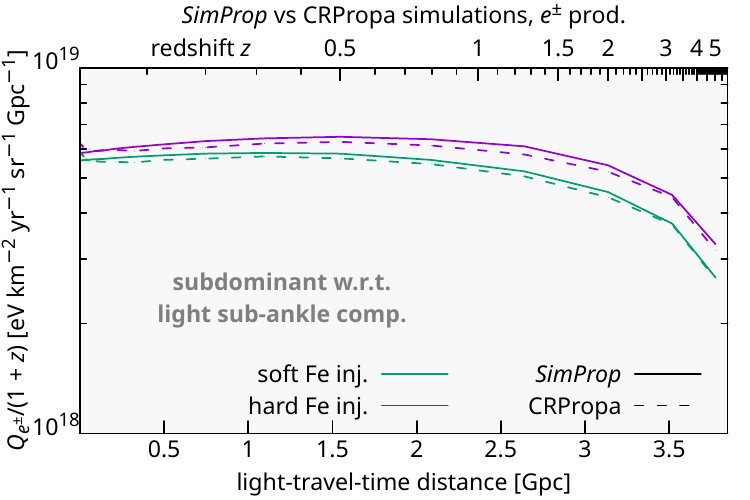}\\
	\includegraphics[width=3in]{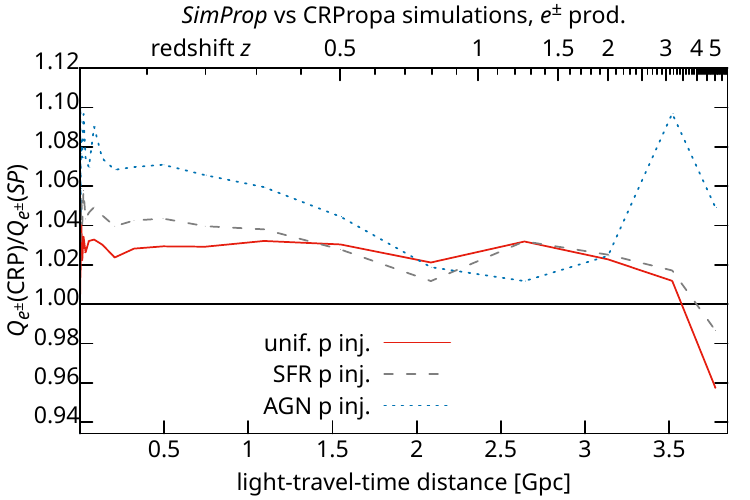}
	\includegraphics[width=3in]{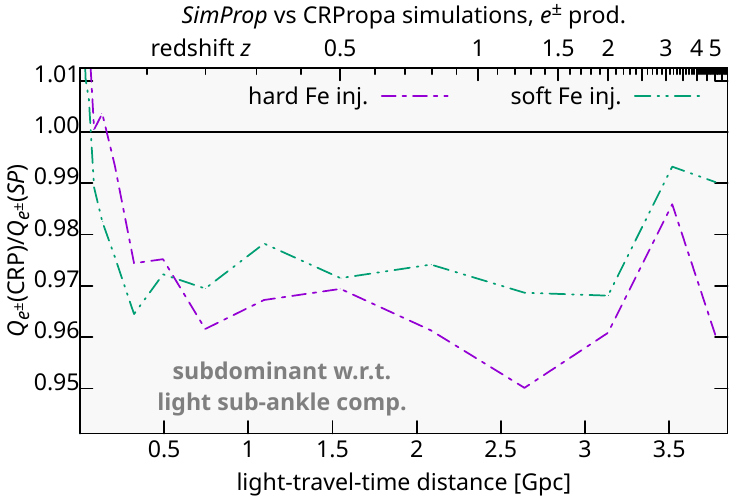}\\
	\includegraphics[width=3in]{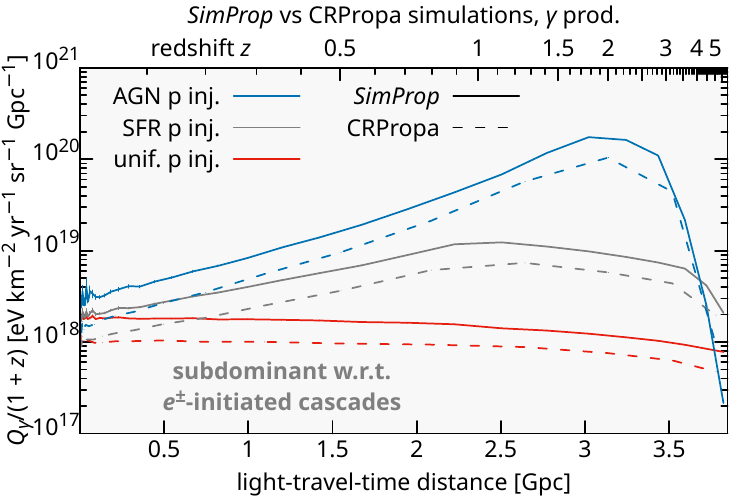}
	\includegraphics[width=3in]{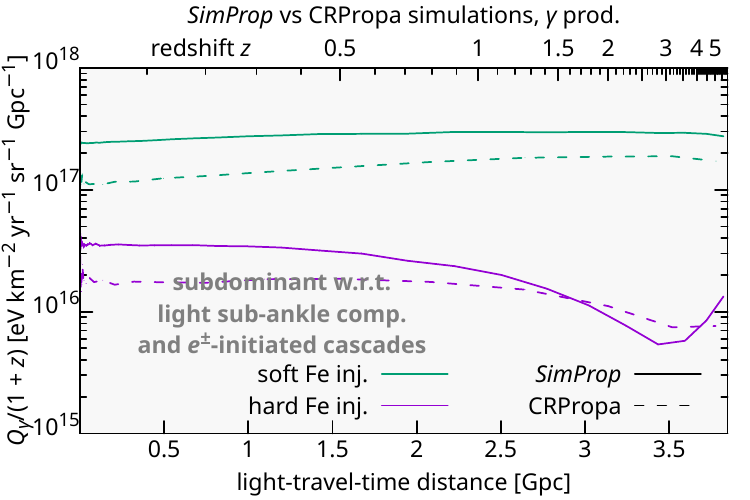}\\
	\includegraphics[width=3in]{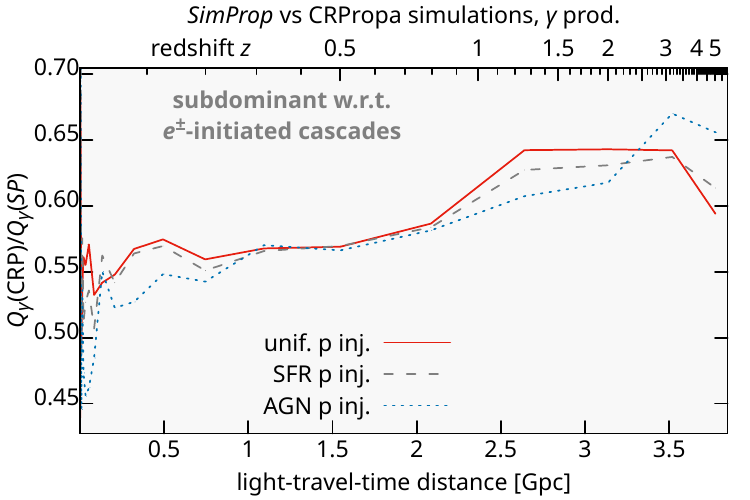}
	\includegraphics[width=3in]{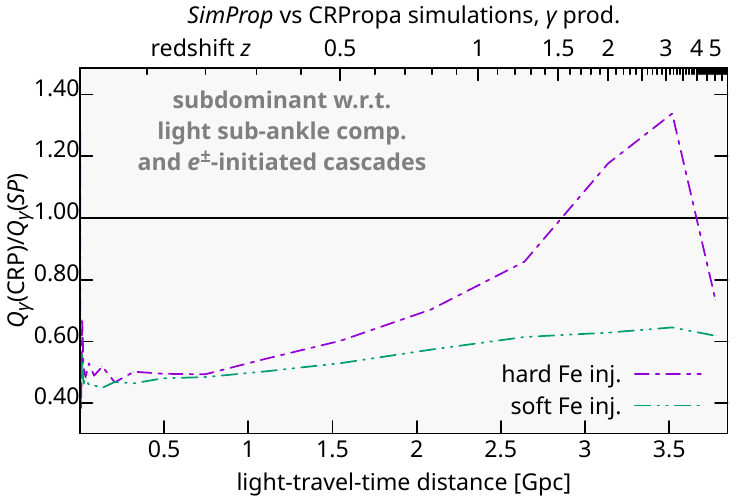}%
	\caption{Cosmogenic $e^\pm$ and $\gamma$ production rates predicted by two different simulation codes,
		and their ratios.  Differences for nitrogen injection are slightly smaller than for iron, and not shown.}
	\label{fig:codes-gammacomp}
\end{figure}

When used with the same EBL and photodisintegration models, \SimProp and CRPropa give very similar results for cosmic-ray fluxes at Earth, as shown in Ref.~\cite{Batista:2015mea}, with discrepancies of $\lesssim 30\%$ for hard nitrogen injection,
$\lesssim 10\%$ for iron or soft nitrogen injection, and even less for proton injection.  On the other hand, we find that the predicted EeV neutrino fluxes (\figureref{fig:codes-nucomp}) and the densities of cascades initiated by photons (\figureref{fig:codes-gammacomp}) differ by $\sim 50\%$, the former being larger in CRPropa and the latter in \SimProp.\footnote{Coincidentally, \SimProp versions up to v2r3 inclusive included a bug in the pion production branching ratios which caused them to underestimate neutrino fluxes and overestimate photon production rates by a factor of 2, so results from older versions would fortuitously be in better agreement with CRPropa ones.}  For PeV neutrino fluxes, the difference is even larger.  On the other hand, predictions of electron/positron production rates (with respect to which photon production rates are subdominant) from the two codes are in agreement to within a few per~cent.

A comparison between neutrino flux predictions of \SimProp and CRPropa and those from Ref.~\cite{Kotera:2010yn} demonstrate that these are in good agreement with CRPropa.
Note that this agreement is only achieved if the same EBL model as in Ref.~\cite{Kotera:2010yn} is used, namely that of Stecker et~al.~(2006)~\cite{Stecker:2005qs,Stecker:2006eh}. More recent models we used in our analysis, and discussed in the \sectionref{sec:EBL_comp}, result in PeV neutrino fluxes several times lower, showing the claim in Ref.~\cite{Kotera:2010yn} that the choice of EBL model is irrelevant to be incorrect.

\begin{figure}
	\centering
	\includegraphics[width=3in]{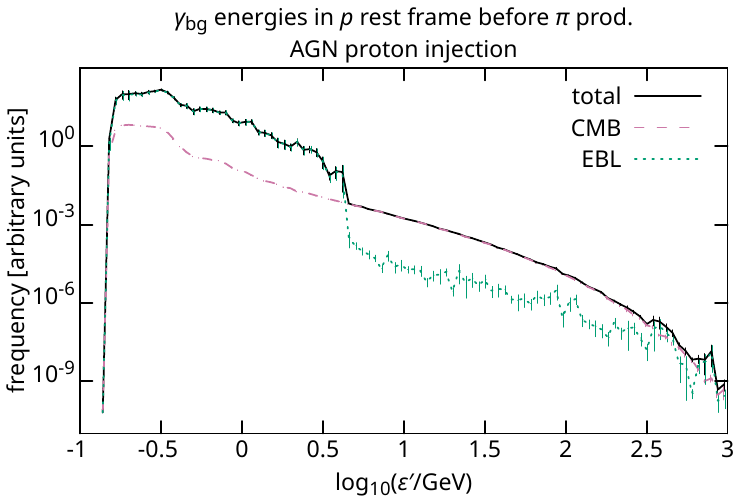}%
	\caption{Distribution of photon energies in the nucleus rest frame in photohadronic interactions during UHECR propagation, computed using a modified \SimProp version, showing that most of the interactions occur near the kinematical threshold.}
	\label{fig:epsprime}
\end{figure}

The main difference between \SimProp and CRPropa relevant to neutrino and photon production is that \SimProp treats all photohadronic interactions as single-pion production, with branching ratios from isospin invariance (i.e.~$\pi^0 : \pi^\pm = 2 : 1$), and with the angular distribution of the outgoing pion assumed isotropic in the CoM~frame, whereas CRPropa also includes channels producing several and/or heavier hadrons, with branching ratios and angular distributions taken from SOPHIA~\cite{Mucke:1999yb}.  At interaction energies close to the threshold, which are the most frequent ones due to the steeply falling cosmic-ray and background photon spectra (see \figureref{fig:epsprime}), SOPHIA predicts a lower $\pi^0 : \pi^\pm$ ratio than isospin invariance, and hence fewer photons and more neutrinos.  Another difference relevant for neutrino fluxes at $\lesssim 10^{17}$~eV, is that \SimProp uses a 2D interpolation for the EBL spectrum and evolution, whereas CRPropa approximates it as the product of two 1D functions for energy and redshift.

\begin{figure}
	\centering
	\includegraphics[width=3in]{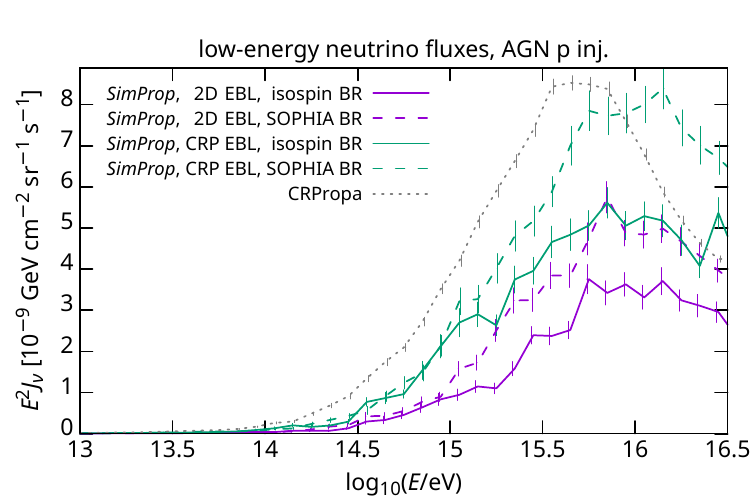}
	\includegraphics[width=3in]{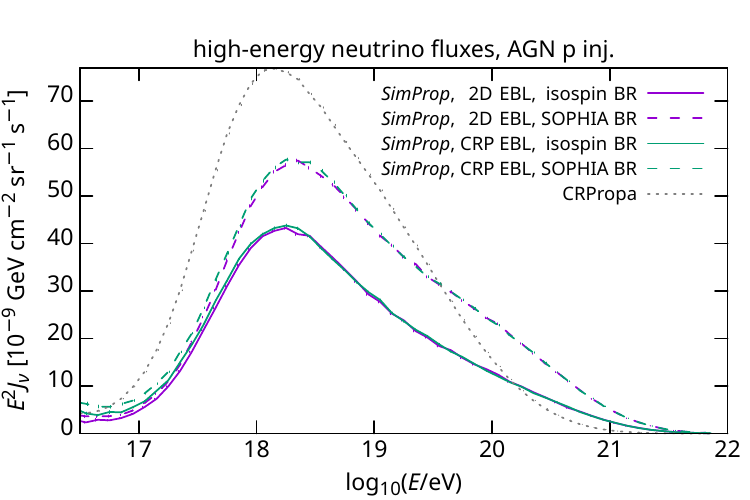}%
	\caption{Effects of approximations in the EBL redshift scaling and in pion production on predicted neutrino fluxes.
	The publicly released version of \SimProp corresponds to the solid black line.
	Note that the $y$-axis is not logarithmic in order to highlight small differences.}
	\label{fig:test}
\end{figure}

Using specially modified versions of \SimProp implementing the same EBL scaling scheme as CRPropa and the same $\pi^0 : \pi^\pm$ ratio as predicted by SOPHIA, we determined that these two differences largely explain the discrepancies in predicted neutrino and photon productions.  The difference between EBL treatments is slightly more important than that between pion production branching ratios for low-energy neutrino fluxes, but is negligible for high-energy neutrinos, as shown in \figureref{fig:test}.
A difference in the shape of the neutrino spectra remains, \SimProp having a thicker high-energy tail and CRPropa a higher low-energy shoulder, presumably due to the remaining kinematical approximations used in \SimProp.  \SimProp developers are considering implementing more realistic pion production models in future \SimProp versions.

	\section{Discussion}\label{sec:discussion}

Our results have immediate implications for phenomenological studies. The level of uncertainty of the physical and astrophysical ingredients to be included in the calculation of the propagation of the UHECRs through extragalactic space, as well as the approximations made in the computations, are known to influence the interpretation of UHECR data~\cite{Batista:2015mea,Aab:2016zth}. The guaranteed neutrino and photon fluxes produced in the propagation are then studied in this work in order to quantify how the uncertainties in the UHECRs propagate to the secondary messengers.

In Refs.~\cite{Romero-Wolf:2017xqe,AlvesBatista:2018zui,Heinze:2019jou} the authors have attempted to fit the spectrum and composition measured by the Pierre Auger Observatory in terms of simple source models, and based on that, to compute fluxes of cosmogenic neutrinos. As shown in our previous work~\cite{Batista:2015mea} and later in~\cite{Aab:2016zth}, the fit is fairly sensitive to a number of assumptions including simulation codes, photodisintegration cross sections, EBL models, etc. As a consequence, these uncertainties may propagate to the fluxes of secondary particles, thereby yielding considerably large errors.

It is interesting to stress that characteristics of the primary UHECRs can magnify or hide the differences in the secondary neutrino or photon spectra. The UHECR data, if a simple astrophysical source model is used, seem to favour scenarios with hard spectral indices and low rigidity cutoffs.
In these scenarios, cosmogenic neutrino fluxes are more sensitive to details of UHECR propagation than in the ones with soft spectral indices and high rigidity cutoffs,  making our study very prominent in the context of a global view on UHECR, neutrino, and photon data.  On the other hand, in such scenarios the high-energy neutrino fluxes are much reduced, and our computed low-energy neutrino fluxes from nuclei do not take into account the presence of a light cosmic-ray component below the ankle, which presumably results in a low-energy neutrino flux with respect to which the one produced by the highest-energy nuclei would be subdominant.


We showed that the effects of details of UHECR propagation on electron/positron production rates are minor; those on photon production rates are larger, but the latter's overall contributions to electromagnetic cascades are much smaller, so their uncertainties are less important.  In most cases, the differences between models in electron/positron production and in photon production have different signs, further reducing the net effect on the overall cascade density.
However, in this work we did not study the development of the cascades, which may be sensitive to differences between models even in cases not relevant to UHECRs.
For example, the universal radio background (URB) only affects UHECR propagation at extremely high energies ($E \gtrsim 10^{22}~\eV$), thus being virtually negligible for our study, but has considerable effects on the development of electromagnetic cascades. Discrepancies exist across URB models. For instance, the model by Protheroe \& Biermann~\cite{Protheroe:1996si} is implemented in CRPropa and used for cascade propagation. However, recent measurements by the ARCADE-2~\cite{Fixsen:2009xn} suggest that the overall intensity of this background exceeds the expectations. Therefore, a more detailed investigation of uncertainties of URB models on particle propagation, in particular ultra-high-energy photons, is needed.

In Ref.~\cite{Soriano:2018lly} a thorough study of the photodisintegration of $^4\text{He}$ in the CMB was conducted. Using up-to-date photonuclear data, the authors show that the effective survival probability of helium would increase significantly. As a consequence, the flux of 63 EeV helium nuclei from a source located at 3.5 Mpc would be 35\% and 42\% larger than predicted by CRPropa and by \SimProp, respectively. While the parametrisation from Ref.~\cite{Soriano:2018lly} has not yet been implemented in either codes, its consequences can be qualitatively understood -- cosmogenic fluxes would decrease because mean free paths would be larger. 

We found that cosmogenic neutrino fluxes are strongly dependent on branching ratios of photohadronic interactions at relatively low $\epsilon'$ (photon energy in the nucleus rest frame). In most studies these are either computed assuming isospin invariance or with the SOPHIA code~\cite{Mucke:1999yb}.  The former overestimate the $p:n$ ratio, and hence the secondary $\gamma:\nu$ ratio, at low $\epsilon'$. The latter are in very good agreement with the data at $\epsilon'\lesssim 1$~GeV, but at higher energies the data available are sparse and seemingly in tension with SOPHIA predictions ($p:n\approx 3.8$ in the data, 2.2 in SOPHIA), though with large uncertainties.  Fortunately such energies seldom occur in UHECR interactions in the intergalactic space, despite the fact that this might not necessarily be the case in interactions with denser, higher-energy radiation fields in the immediate vicinity of the sources. In addition, approximations of the photomeson production by nuclei (meaning the scaling of the cross section with the mass $A$ and the superposition model) could have a direct impact on the expected cosmogenic fluxes. However, the effect on these fluxes would only be noticeable if the typical energies of the UHECRs were much larger than the ones found through fits of the UHECR spectrum and composition. On the other hand, the effects of these approximations become, also in this case, more relevant in interactions with other radiation fields surrounding the sources. 

Due to the UHECR horizon, UHECR data do not provide any information about the redshift evolution of the EBL up to distant regions of the Universe. Conversely, in principle, cosmogenic neutrino fluxes could be used as a diagnostic tool for understanding astrophysical inputs; in practice, the fluxes are much more sensitive to the UHECR mass composition and cosmological evolution of the sources than to details of the EBL. In fact, for the cosmogenic neutrino flux, the source evolution and fractions of each nuclear species are strongly correlated and hard to disentangle~\cite{Moller:2018isk}. Therefore, using cosmogenic neutrino data to infer information about the EBL will not be possible unless and until these quantities are known with much better precision than at present.

Other sources of uncertainties which have not been investigates in this work are the distribution of sources and magnetic fields. In Ref.~\cite{Mollerach:2013dza} the authors argue that diffusion of UHECRs from nearby sources could affect the shape of the spectrum. On the other hand, in realistic magnetic field distributions derived from cosmological magnetohydrodynamical simulations, this effect may not be relevant for nearly homogeneous source distributions~\cite{Batista:2014xza}. Fits of the Pierre Auger Observatory data taking into account EGMF deflections and the local extragalactic matter distribution were presented in Refs.~\cite{Wittkowski:2017okb,Eichmann:2017iyr}. The authors show that the inclusion of magnetic fields and specific source distributions changes the fit results. In addition, this could also affect the expected cosmogenic neutrino and photon fluxes.
A first effort in this direction was done in~\cite{Wittkowski:2018giy}, who computed the cosmogenic neutrino fluxes in light of the best-fit results of Ref.~\cite{Wittkowski:2017okb}. They show that the inclusion of EGMFs enhances the expected fluxes by a factor of a few at $E \sim 1~\text{EeV}$, and by more than two orders of magnitude at $E \gtrsim 10~\text{EeV}$, making it more likely for future UHE neutrino experiments to detect these particles. From that work, however, it remains unclear whether these enhanced cosmogenic neutrino fluxes are due to effects of the EGMF or to the differences in the best-fit source composition, spectral index, and maximum rigidity. In addition, given the large uncertainties across EGMF models (see e.g.~\cite{AlvesBatista:2017vob}), it is at the moment neither possible to provide estimates of how much these fields would impact the production of secondary particles nor whether these uncertainties dominate over the ones studied in Sec.~\ref{sec:comparisons}.

Whereas cosmogenic fluxes are useful diagnostic tools in the study of UHECR source classes, they can also be backgrounds for the detection of individual sources of high-energy gamma rays and neutrinos. Therefore, it is essential to understand the uncertainties in their production as they will have a direct impact on the prospects for detecting point-like sources with future neutrino and gamma-ray telescopes.

	\section{Conclusions}\label{sec:conclusions}
In this paper we have extended our previous work \cite{Batista:2015mea} to analyse the impact of various uncertainties on the production rates of secondary particles, namely, neutrinos, photons, and electron--positron pairs, for protons and heavier UHECR primaries. In particular, we have considered the following sources of uncertainties: cosmological parameters, propagation codes (CRPropa and \SimProp), photodisintegration cross sections, and EBL spectrum.

We have found that the choice of cosmological parameter values has practically no impact on any of the observables at Earth, making it fully justifiable to neglect their uncertainties as commonly done in UHECR propagation studies.

The neutrino fluxes computed with CRPropa and \SimProp differ significantly: the predictions by CRPropa exceed those by \SimProp by over 50\% at EeV energies, and by several times at PeV energies for proton injection. 
We found this to be due to an approximation in the CRPropa implementation of the EBL redshift evolution which results in overestimated PeV neutrino fluxes and an approximation in the \SimProp implementation of pion production resulting in underestimated neutrino fluxes (and overestimated photon production rates) at all energies.  We are considering using more accurate implementations in future versions of the codes.

The choice of photodisintegration model has no noticeable effect on predicted neutrino fluxes, except for beta-decay neutrinos at energies below $10^{17}$~eV, which in any realistic case are subdominant with respect to EBL-produced pion-decay neutrinos. Conversely, different models for the extragalactic background light can affect the neutrino spectrum by several tens of percent. Therefore, we identify EBL models as one of the main sources of uncertainties when computing expected ultra-high-energy neutrino fluxes for a given injection model.

We also found that the production rates of electrons and positrons are not very sensitive to details of UHECR propagation. There are sizeable differences in production rates of photons among propagation models, but these are very subdominant with respect to those of electrons and positrons.  Possible uncertainties on the development of cascades initiated by these particles are outside the scope of this work.

	\acknowledgments
The work by RAB is supported by grant \#2017/12828-4, São Paulo Research Foundation (FAPESP). The work by AdM is supported by the IISN project 4.4502.16. The work by DB and AvV has been supported by the European Research Council (ERC) under the European Union’s Horizon 2020 research and innovation programme (Grant No. 646623). AvV acknowledges financial support from the NWO Astroparticle Physics grant WARP.

	\bibliographystyle{JHEP}
	\bibliography{main}
\end{document}